\pdfoutput=1
\documentclass[useAMS,amssymb,usenatbib]{mn2e}
\usepackage{epsfig}
\usepackage{color}
\usepackage{graphicx}
\usepackage{tabularx}
\usepackage{longtable}
\usepackage{hyperref}           
\usepackage[normalem]{ulem}

\newcommand{\Ms}{\mathrm{M}_\odot}

\newcommand{\vm}{\mathrm{v_{max}}}   
\newcommand{\rmax}{\mathrm{r_{max}}}   

\makeatletter
\newenvironment{chapquote}[2][2em]
  {\setlength{\@tempdima}{#1}%
   \def\chapquote@author{#2}%
   \parshape 1 \@tempdima \dimexpr\textwidth-2\@tempdima\relax%
   \itshape}
  {\par\normalfont\hfill--\ \chapquote@author\hspace*{\@tempdima}\par\smallskip}
\makeatother

\title[The chosen few]{The chosen few: the low mass halos that host
  faint galaxies}

\author[Sawala et al.]  {\parbox{\textwidth}{Till
    Sawala$^{1}$\thanks{E-mail: \texttt{till.sawala@durham.ac.uk}},
    Carlos~S.~Frenk$^1$, Azadeh Fattahi$^2$, Julio~F.~Navarro$^2$, Tom
    Theuns$^{1,3}$, Richard~G.~Bower$^1$, Robert~A.~Crain$^4$,
    Michelle Furlong$^1$, Adrian Jenkins$^1$, Matthieu Schaller$^1$
    and Joop
    Schaye$^4$}\vspace{0.4cm}\\
\parbox{\textwidth}{
$^{1}$Institute for Computational Cosmology, Department of Physics, University of Durham, South Road, Durham DH13LE, UK \\ 
$^{2}$Department of Physics and Astronomy, University of Victoria,
3800 Finnerty Road, Victoria, British Columbia V8P 5C2, Canada\\ 
$^{3}$Department of  Physics, University of Antwerp, Campus Groenenborger, Groenenborgerlaan 171, B-2020 Antwerp, Belgium \\
$^{4}$Leiden Observatory, Leiden University, Postbus  9513, 2300 RA Leiden, The Netherlands \\ 
 }}

\begin{document}

\date{Accepted 2014 ***. Received 2014 ***; in original  form 2014}

\pagerange{\pageref{firstpage}--\pageref{lastpage}} \pubyear{2014}

\maketitle

\label{firstpage}

\begin{abstract}
  Since reionization prevents star formation in most halos less
  massive than $3\times 10^9 \Ms$, dwarf galaxies only populate a
  fraction of existing dark matter halos. We use hydrodynamic
  cosmological simulations of the Local Group to study the
  discriminating factors for galaxy formation in the early Universe
  and connect them to the present-day properties of galaxies and
  halos. A combination of selection effects related to reionization,
  and the subsequent evolution of halos in different environments,
  introduces strong biases between the population of halos that host
  dwarf galaxies, and the total halo population. Halos that host
  galaxies formed earlier and are more concentrated. In addition,
  halos more affected by tidal stripping are more likely to host a
  galaxy for a given mass or maximum circular velocity, $\vm$,
  today. Consequently, satellite halos are populated more frequently
  than field halos, and satellite halos of $10^8-10^9\Ms$ or $\vm$ of
  $12-20$~kms$^{-1}$, similar to the Local Group dwarf spheroidals,
  have experienced a greater than average reduction in both mass and
  $\vm$ after infall. They are on closer, more radial orbits with
  higher infall velocities and earlier infall times. Together, these
  effects make dwarf galaxies highly biased tracers of the underlying
  dark matter distribution.
\end{abstract}


\begin{chapquote}[30pt]{Bertolt Brecht, {\it The Threepenny Opera}}
\noindent ``There are  some who are in darkness \\
And the others are in light \\
And you see the ones in brightness \\
Those in darkness drop from sight'' \\
\end{chapquote}

\section{Introduction}\label{sec:introduction}
The Local Group dwarf galaxies and their halos are often described as
ideal test cases for the impact of cosmological models on small scale
structures. It has, of course, long been recognised that astrophysical
processes such as reionization can suppress star formation in low mass
halos \citep[e.g.][]{Efstathiou-1992, Bullock-2000, Benson-2002,
  Somerville-2002}. It should therefore perhaps come as no surprise
that the total number of substructures predicted in a given
cosmological model can be far greater than the number of observable
galaxies. However, as reionization occurs in the early universe when
halos are at a fraction of their final mass, it is also expected that
the properties of Local Group halos and galaxies today are only
loosely related to the conditions which separated those progenitor
halos that were able to form stars from those that have remained dark.

In this paper we examine how the sparse sampling of halos by galaxies
in the presence of an ionizing background, and the subsequent
evolution of halos and galaxies in a Local Group environment, can lead
to a population of halos hosting faint galaxies that is very different
from the total halo population.

Previous simulations have studied the impact of reionization both for
individual dwarf galaxies and for satellites of Milky Way sized
halos. \cite{Okamoto-2008} and \cite{Okamoto-2009} have found that
reionization can remove most of the baryons from halos with low
maximum circular velocity, $\vm$ (the circular velocity measured at
the radius where $\rm{v_{circ}} = \sqrt{GM(<r) / r}$ is maximal), and
showed that dark satellite halos are expected around the Milky Way
with $\vm$ up to 25~kms$^{-1}$. They concluded that $\vm \sim
12$~kms$^{-1}$ at the time of reionization ($z=8$ in their model)
sharply separates dark and luminous halos. \cite{Nickerson-2011} have
also performed a simulation of a Milky-Way mass halo and its
satellites, but found the peak mass, $\rm{M_{peak}}$, which satellites
reach during their evolution to be the discriminating factor, with
halos of $\rm{M_{peak}} < 2\times10^9\Ms$ losing most of their gas to
reionization and failing to form stars. \cite{Shen-2013} simulated
galaxy formation in seven halos of mass between $4.4 \times10^8$ and
$3.6\times10^{10}\Ms$, and found that the three halos with a peak
value of $\vm < 16$ kms$^{-1}$ were devoid of stars, and two halos
with peak mass of $1.8$ and $3.3\times 10^9\Ms$ only started star
formation at $z\sim0.5$, long after reionization.

Recently, we used a set of cosmological hydrodynamic simulations of
Local Group like volumes to show that the appearance of ``dark'' halos
that host no galaxy, combined with the reduction in halo mass due to
baryonic processes, can significantly change the expected
stellar-to-total mass relation of dwarf galaxies, and resolve
previously reported discrepancies between observations of individual
dwarf galaxies and the predictions of $\Lambda$CDM derived from
abundance matching (\cite{Sawala-2014}, see also
\cite{Sawala-matter}). In this paper we use the same set of
simulations to study how reionization selects between halos that host
galaxies and those that do not. Following the formation and evolution
of dark and luminous halos throughout cosmic time, we can connect
their properties in the early universe to the observable galaxies and
halos of the present day.

We also link individual halos from the hydrodynamic simulations to
their counterparts in dark matter only (DMO) simulations of the same
volumes. This allows us to study the intrinsic properties of halos
which determine galaxy formation, and distinguish them from the impact
that the baryonic processes associated with galaxy formation can have
on individual halos. Finally, by comparing the populations of dark and
luminous halos in the simulated Local Group volumes, we deduce the
biases which can arise when observed galaxies are assumed to be
representative of the underlying population of dark matter halos.

This paper is organised as follows. In Section~\ref{sec:methods}, we
briefly describe the simulations and the astrophysical assumptions
made. By comparing simulations with and without reionization, we
discuss the impact of reionization in
Section~\ref{sec:reionization}. Section~\ref{sec:timing} describes
galaxy formation in the presence of reionization. In
Section~\ref{sec:assembly}, we explain how the assembly histories
separate those halos which form stars from those that remain dark. In
Section~\ref{sec:formation-redshift} we discuss the resulting bias in
formation redshifts, and in Section~\ref{sec:velocity-mass} we show
the difference in concentration between luminous and non-luminous
halos. In Section \ref{sec:bias}, we examine how the subsequent
evolution in different environments within the Local Group creates
further biases between luminous and non-luminous halos: satellites
halos that are luminous have higher infall redshifts and more radial
orbits with higher infall velocities (Section~\ref{sec:satellites}),
and the distributions of luminous and non-luminous halos differ within
the Local Group (Section~\ref{sec:environment}). We discuss some
implications of our findings and conclude with a summary in
Section~\ref{sec:summary}.

\begin{figure*}
  \begin{center}
\vspace{-.15in}
 \fbox{\includegraphics*[trim = 0mm 0mm 0mm 0mm , clip, width =
 .472\textwidth]{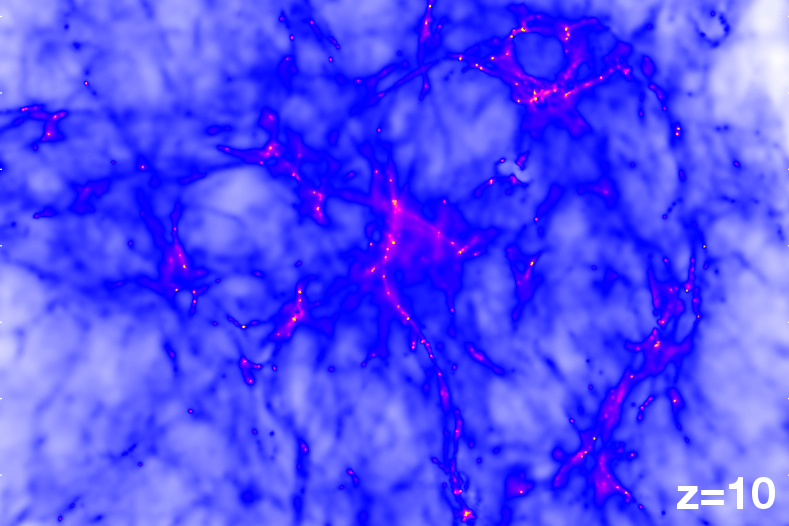}}
  \fbox{\includegraphics*[trim =0mm 0mm 0mm 0mm , clip, width =
 .472\textwidth]{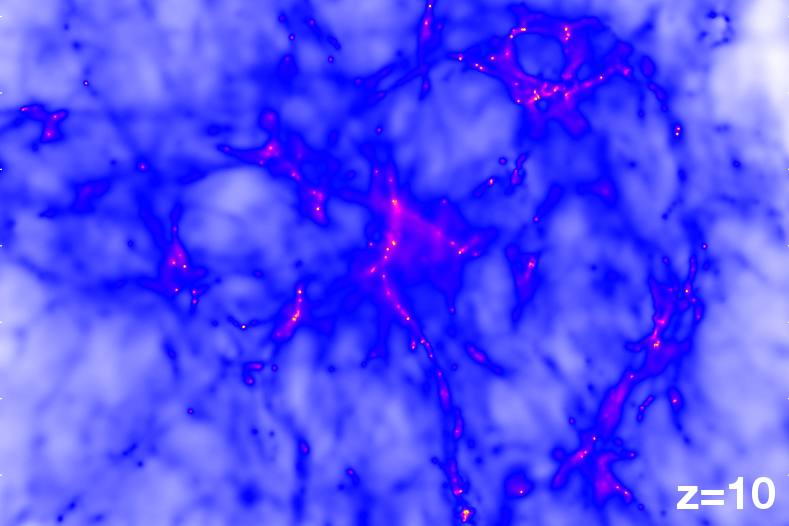}} \\
 \fbox{\includegraphics*[trim = 0mm 0mm 0mm 0mm , clip, width =
 .472\textwidth]{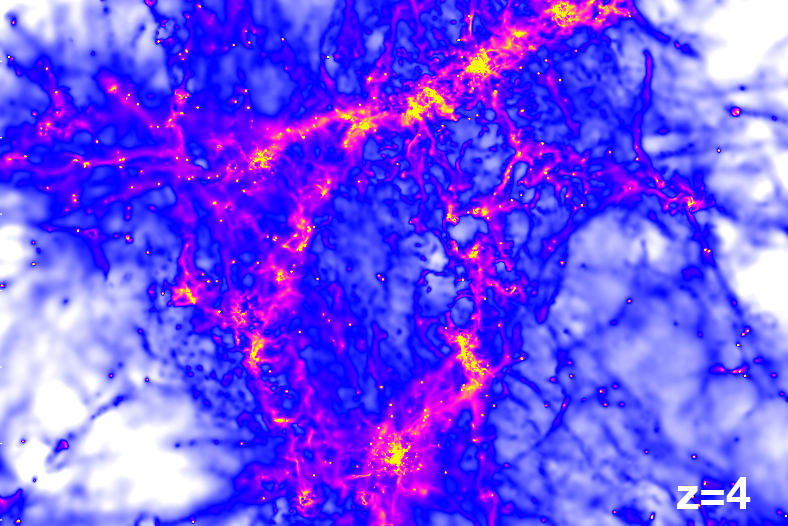}} 
 \fbox{\includegraphics*[trim = 0mm 0mm 0mm 0mm , clip, width =
 .472\textwidth]{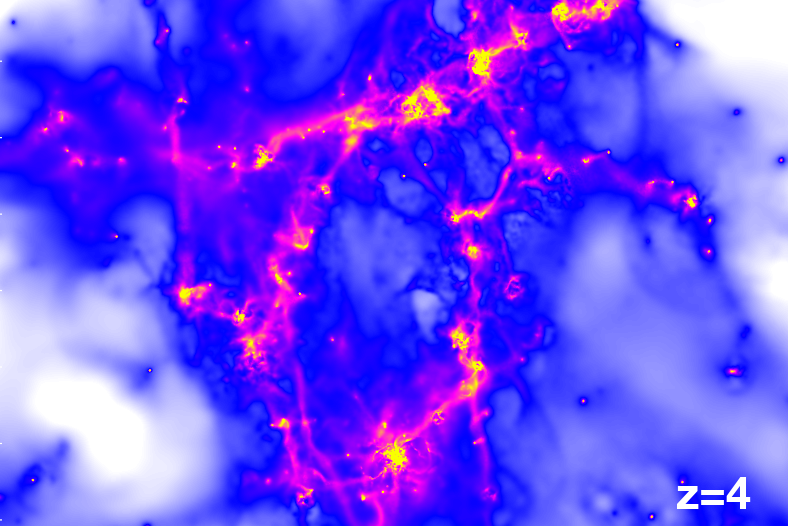}} \\
 \fbox{\includegraphics*[trim = 0mm 0mm 0mm 0mm , clip, width =
 .472\textwidth]{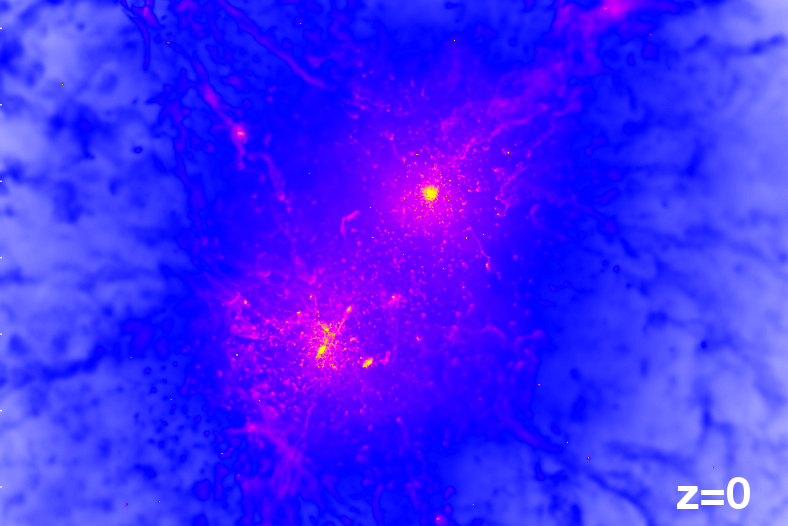}}
 \fbox{\includegraphics*[trim = 0mm 0mm 0mm 0mm , clip, width =
.472\textwidth]{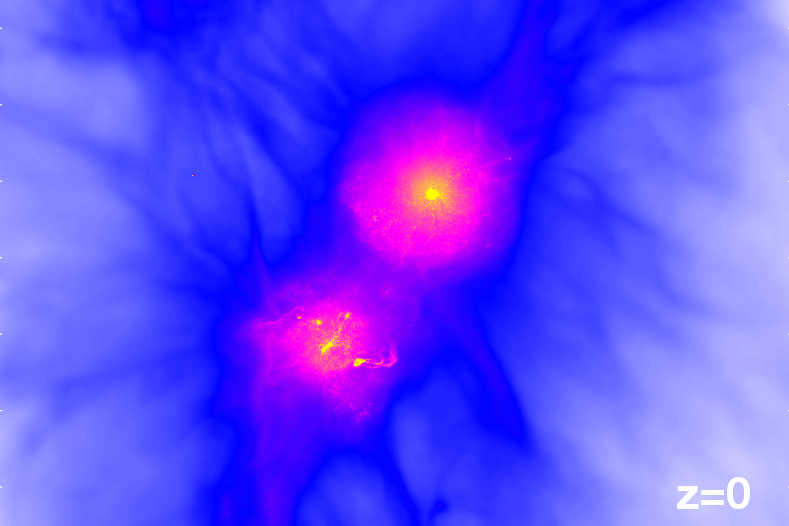}}  \\
 \fbox{\includegraphics*[trim = 0mm 0mm 0mm 0mm , clip, width =
.472\textwidth]{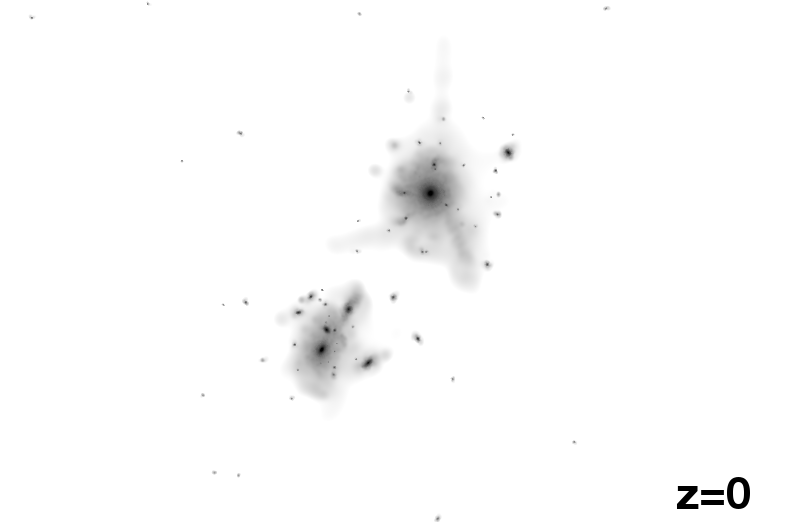}} 
 \fbox{\includegraphics*[trim = 0mm 0mm 0mm 0mm , clip, width =
.472\textwidth]{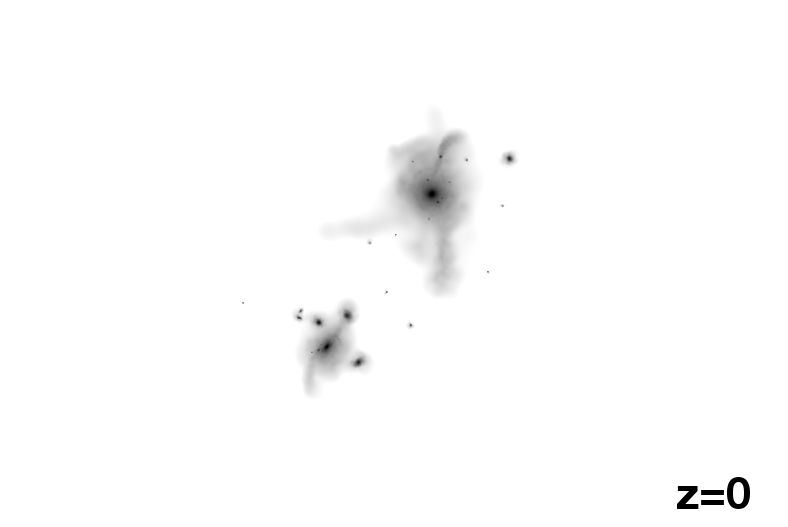}} \\
\vspace{-.15in}

 \end{center}
 \caption{Gas density distribution (top three rows) and stellar
   density distribution (bottom row, only halos with $m_\star >
   4\times10^6\Ms$) in the L2 simulation without reionization (left)
   and the same volume with reionization (right). The region shown
   measures 3$\times$~2Mpc$^2$ at $z=0$ and is magnified proportional
   to $a(z)$ at higher redshifts. Following reionization at $z=11.5$,
   gas is removed from low mass halos, and reduced cooling results in
   a less fragmented IGM. The impact of reionization is first
   noticeable in the low-density regions, which are already less
   structured at $z=10$, just after reionization. At later times, the
   differences increase, and the number of collapsed and eventually
   star-forming halos is much lower in the simulation that includes
   reionization.}
  \label{fig:baryon-density}
\end{figure*}

\section{Methods}\label{sec:methods}
The simulations used in this paper have previously been described in
\cite{Sawala-2014}. We reproduce here the main aspects of their
initial conditions and astrophysical model.

We resimulate 12 cosmological volumes as ``zoom'' simulations
extracted frome the {\sc Dove} simulation, a $100^3$Mpc$^3$ N-Body
simulation based on the WMAP-7 cosmology. We require that each volume
should contain two halos of mass in the range
$\left(5\times10^{11}-2.5~\times 10^{12}\right)\Ms$ separated by $800
\pm 200$ kpc, approaching with radial velocity of $\left(0-250\right)$
kms$^{-1}$ and with tangential velocity below $100$ kms$^{-1}$ in an
environment with an unperturbed Hubble flow out to 4 Mpc. The high
resolution initial conditions were created using second-order
Lagrangian perturbation theory, as described by
\cite{Jenkins-2010}. The selection of Local Group environments is
discussed in more detail in Fattahi et al. (2014, in prep).

\begin{table}
  \caption{Numerical parameters of the simulations} 
\centering 
\begin{tabular}{l l c c c} 
\hline
\hline 
& & \multicolumn{2}{c}{Particle Masses} & Max Softening\\
Label & Type & DM $[\Ms]$ & Gas $[\Ms]$ & [pc] \\ [0.5ex] 
\hline 
L1 & hydro & $5.0\times 10^{4}$ &  $1.0\times 10^{4}$ & 94 \\ 
L1 & DMO & $6.0\times 10^{4}$  & -- & 94 \\
L2 & hydro & $5.9\times 10^{5}$ &  $1.3\times 10^{5}$ & 216 \\ 
L2 & DMO & $7.2\times 10^{5}$  & -- & 216 \\
L3 & hydro & $7.3\times 10^{6}$ &  $1.5\times 10^{6}$ & 500 \\ 
L3 & DMO & $8.8\times 10^{6}$  & -- & 500 \\ [1ex] 
\hline 
\vspace{-.3cm}
\end{tabular}
\label{table:params} 
\end{table}
The simulations were performed using a pressure-entropy variant
\citep{Hopkins-2012} of the Tree-PM SPH code {\sc P-Gadget3}
\citep{Springel-2005}, described in Dalla Vecchia et al. 2014 (in
prep.). The subgrid physics model is that of the {\it Evolution and
  Assembly of GaLaxies and their Environments} project ({\sc Eagle},
Schaye et al. 2014 in prep., Crain et al. 2014 in prep.). It includes
metal-dependent radiative cooling \citep{Wiersma-2009} and
photo-heating in the presence of UV and X-ray backgrounds, and the
cosmic microwave background (CMB).

Prior to reionization, net cooling rates are computed from the CMB and
from a UV and X-ray background that follows the $z=9$ model of
\cite{Haardt-2001} with a 1~Ryd high-energy cutoff. To account for the
temperature boost due to radiative transfer and non-equilibrium
effects over the optically thin limit assumed in our simulations
\citep{Abel-1999}, we inject 2~eV per hydrogen and helium atom. We
assume that hydrogen reionizes instantaneously at $z=11.5$
\citep{Planck-2013}, while the redshift dependence of helium
reionization is modelled as a Gaussian centred at $z=3.5$
\citep{Theuns-2002} with $\sigma(z) = 0.5$. As shown by
\cite{Wiersma-2009b} and \cite{Rollinde-2013}, the resulting evolution
of the temperature-density relation is consistent with measurements of
the intergalactic medium \citep{Schaye-2000}.

Star formation follows the formulation of \cite{Schaye-2008} with a
metallicity-dependent threshold \citep{Schaye-2004}. The model
includes stellar evolution \citep{Wiersma-2009b} and stochastic
thermal supernova feedback \citep{Dalla-Vecchia-2012}, as well as
black-hole growth and AGN feedback \citep{Rosas-Guevara-2013,
  Booth-2009, Springel-2005b}.

All simulations were run twice: once with gas and the baryon physics
described above, and once as dark matter only (DMO). In addition, one
volume was also run with the complete hydrodynamic model, but without
reionization. In the DMO simulations the dark matter particle masses
are larger by a factor of $(\Omega_b+\Omega_{DM})/\Omega_{DM}$
relative to the corresponding hydrodynamic simulations. To investigate
the regime of Local Group dwarf galaxies, we use three different
resolution levels labelled L1, L2 and L3, whose parameters are given
in Table~\ref{table:params}. In this work, L3 is only used to test
convergence. The main results for the model that includes reionization
are obtained from five pairs of hydrodynamic and DMO simulations at
resolution L2 and one pair at L1. The simulations without reionization
presented in Section~\ref{sec:reionization} were only run up to L2.

We use a Friends-of-Friends algorithm \citep[FoF;][]{Davis-1985} to
identify overdense structures (FoF-groups), and the {\sc subfind}
algorithm \citep{Springel-Subfind, Dolag-2009} to identify self-bound
substructures within them. As they represent the objects most directly
associated with individual galaxies, we always refer to the self-bound
substructures as ``halos''.  The principal substructure within an
FoF-group contains most of its mass, but satellites may share the same
FoF-group while still residing in separate self-bound
halos. Throughout this paper, we use the term ``satellite'' when we
refer to the satellite halos or galaxies associated with the M31 and
Milky-Way analogues, and ``field'' when we refer to isolated halos.

We analyse our simulations at 128 snapshots, and trace the evolution
of individual halos in both the hydrodynamic and DMO simulations using
merger trees, as described in \cite{Helly-2003} and Qu et al. (2014,
in prep.). The unique IDs of dark matter particles which encode their
positions in the initial conditions allow us to match and compare
individual halos from different simulations of the same volume at the
same resolution.

\section{The impact of reionization}\label{sec:reionization}

From $z=11.5$, the UV background heats the intergalactic medium and
lowers its cooling rate. It can also remove gas from low-mass halos by
photo-evaporation. In Fig.~\ref{fig:baryon-density} we compare the
evolution of the gas density distributions in two simulations of the
same volume and resolution (L2) with and without reionization, as well
as the final stellar density distribution. At $z=10$, shortly after
hydrogen reionization, the main difference is apparent in the
low-density regions. Here, the thermal energy provided by reionization
slows the collapse of small structures which results in a smoother IGM
than in the absence of reionization. By comparison, regions of higher
density which correspond to halos that have already formed before
reionization, are not significantly affected. By $z=4$, the
intergalactic medium has become significantly more fragmented in the
simulation without reionization, with many more low-mass halos now
containing dense gas and forming stars compared to the simulation with
reionization. At $z=0$ it can be seen that while the large-scale
features in both the gas- and stellar density distributions are
similar, in the absence of reionization, the IGM is strongly
fragmented and has collapsed into many small clumps. By contrast, the
IGM in the simulation with reionization has remained much
smoother. Without reionization, the number of halos within 2.5 Mpc
from the LG centre that contain stars is $\sim 700$, compared to only
$\sim 180$ in the same volume with reionization.

\begin{figure}
  \begin{center}
\vspace{-.1in}
  \includegraphics*[trim = 30mm 5mm 20mm 10mm, clip,    width = .46\textwidth]{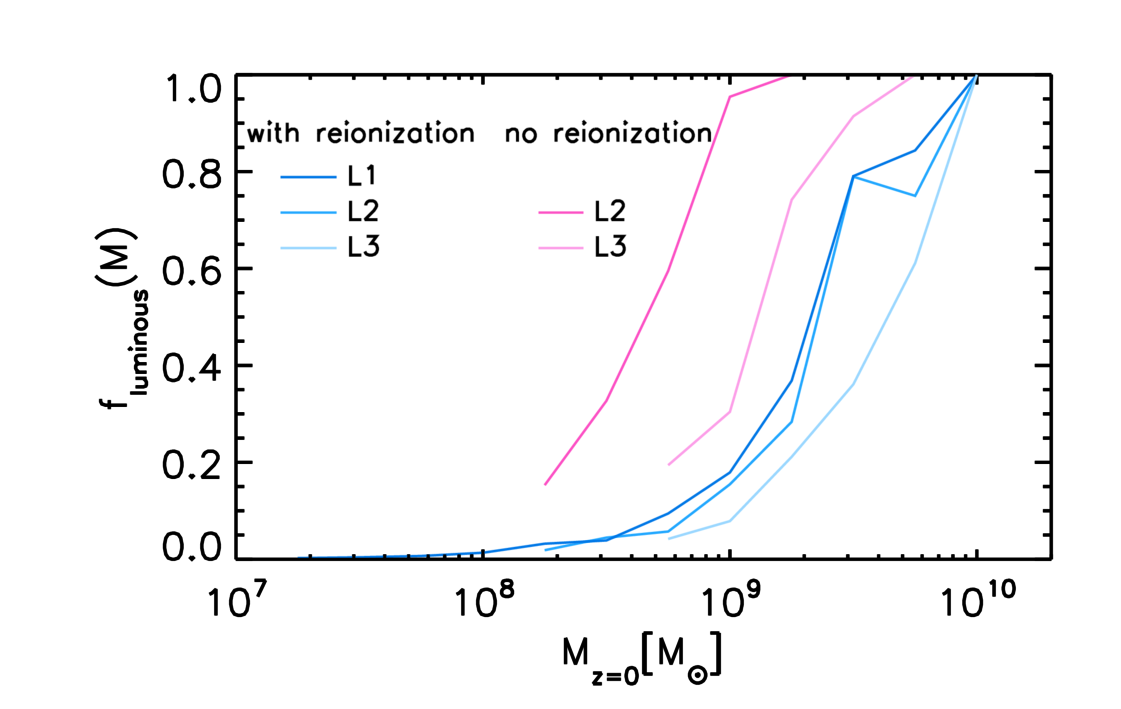}

 \end{center}
\vspace{-.1in}
\caption{Fraction of halos which are luminous at $z=0$ as a function of halo
  mass in simulations with and without reionization for different
  resolutions. When reionization is included, the fraction of luminous
  halos as a function of mass is much reduced. In the simulation
    with reionization, the luminous fraction is converged at L2. By
    contrast, without reionization, the luminous fraction is not
    numerically converged, and would increase further with higher
    resolution.
    \label{fig:luminous-fraction} }
\end{figure}

\begin{figure}
  \begin{center}
\vspace{-.15in}

 \includegraphics*[trim = 30mm 0mm 15mm 10mm, clip,    width =
 .48\textwidth]{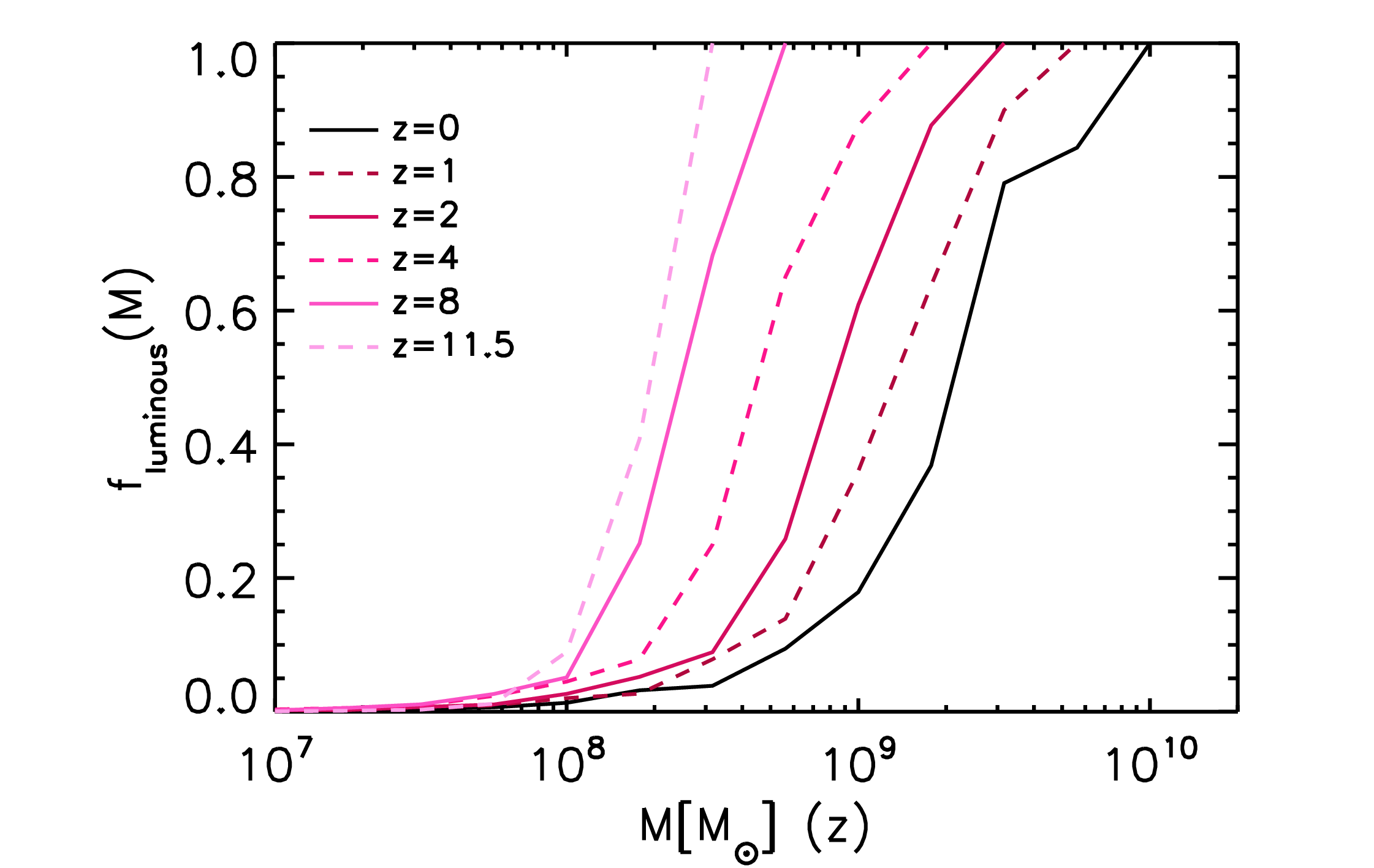}
 \end{center}

\vspace{-.05in}
\caption{Fraction of halos which are luminous as a function of halo mass
    at different redshifts from $z=11.5$ to $z=0$ in the simulation
    with reionization at resolution L1. At any redshift, the fraction
    of luminous halos of mass below $10^8\Ms$ is less than $10\%$, and
    almost no halos below $10^{7.5}\Ms$ contain stars. The mass scale
    that separates luminous from dark halos evolves from $\sim~3\times10^8\Ms$ at $z=11.5$ to $\sim 3\times10^9\Ms$ at $z=0$. 
    \label{fig:evolution-luminousfraction} }
\end{figure}

\begin{figure*}
  \begin{center}
\vspace{-.12in}
\includegraphics*[trim = 20mm 155mm 25mm 0mm, clip, height = 49mm]{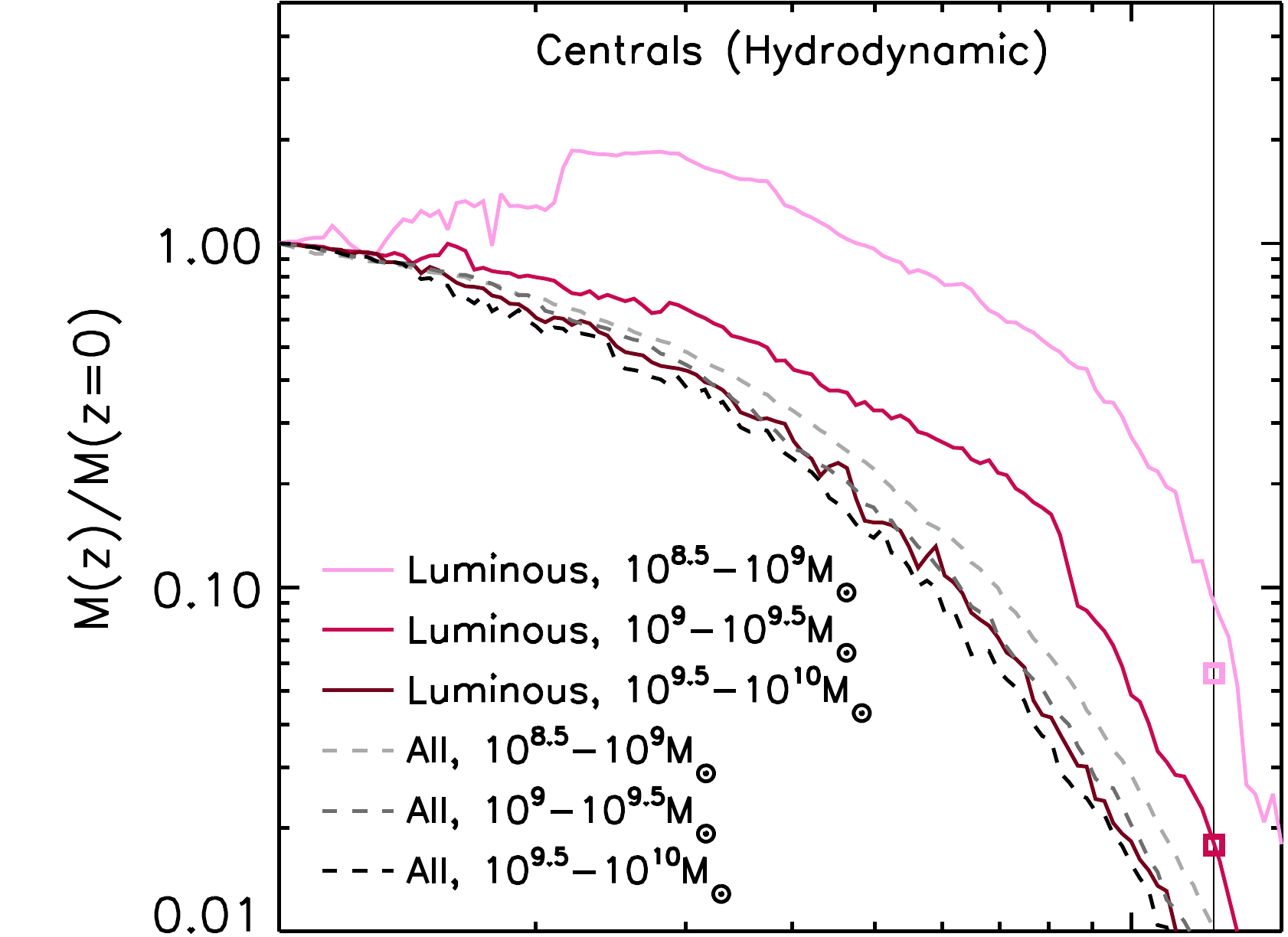}
\includegraphics*[trim = 55mm 155mm 25mm 0mm, clip, height = 49mm]{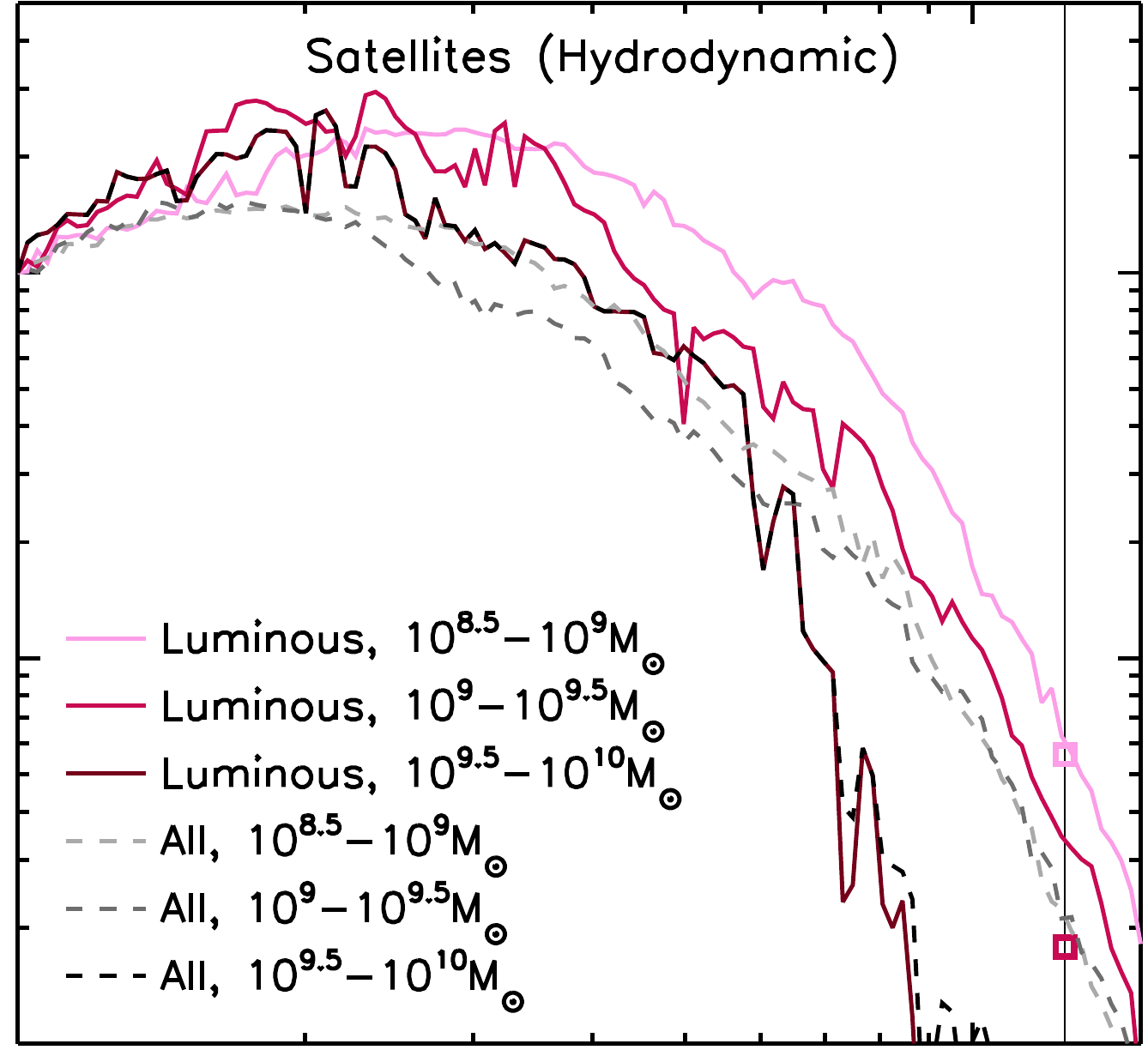} 
\includegraphics*[trim = 55mm 155mm 25mm 0mm, clip, height = 49mm]{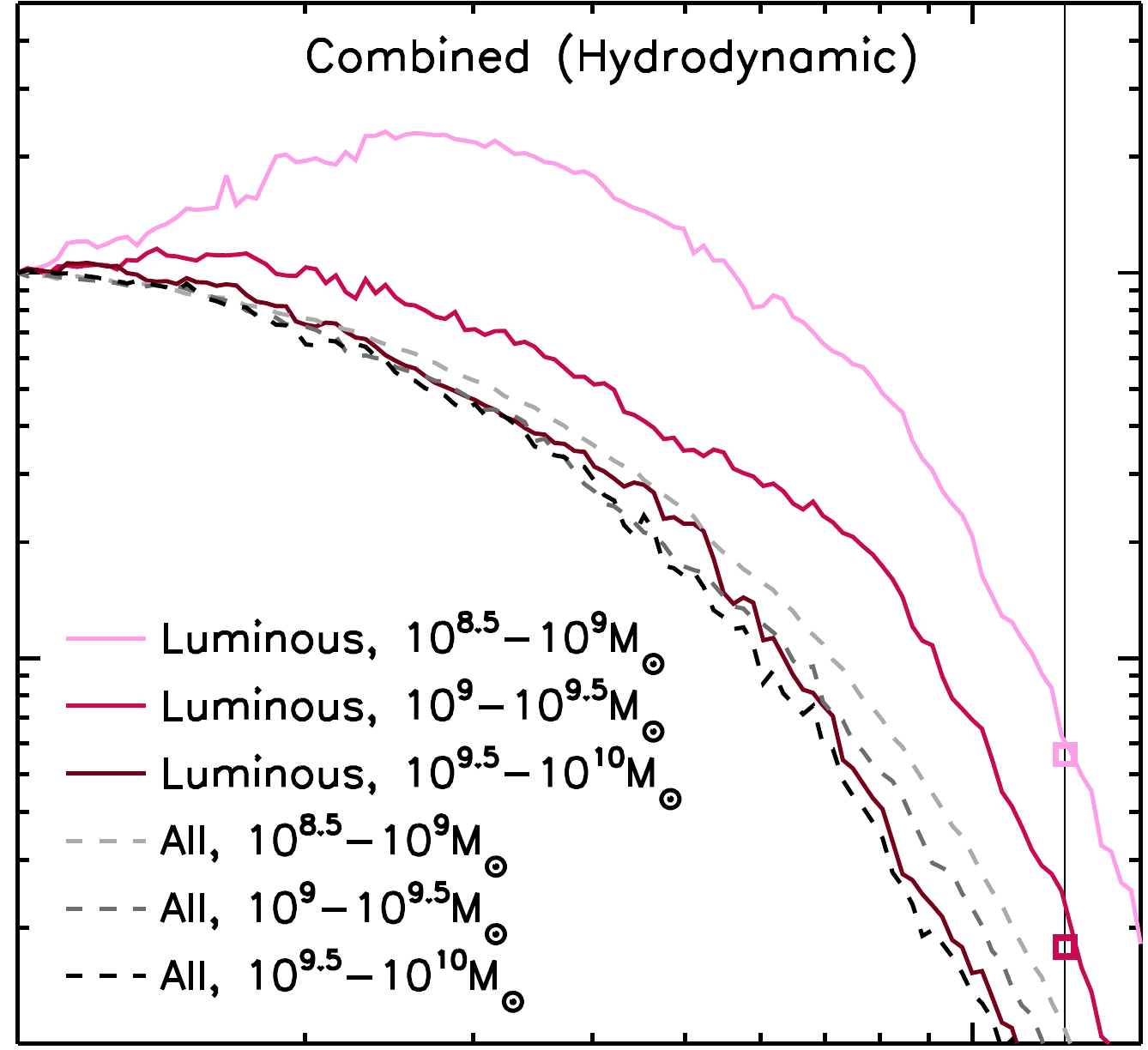}  \\ 
\includegraphics*[trim = 20mm 135mm 25mm 0mm, clip, height = 56.9mm]{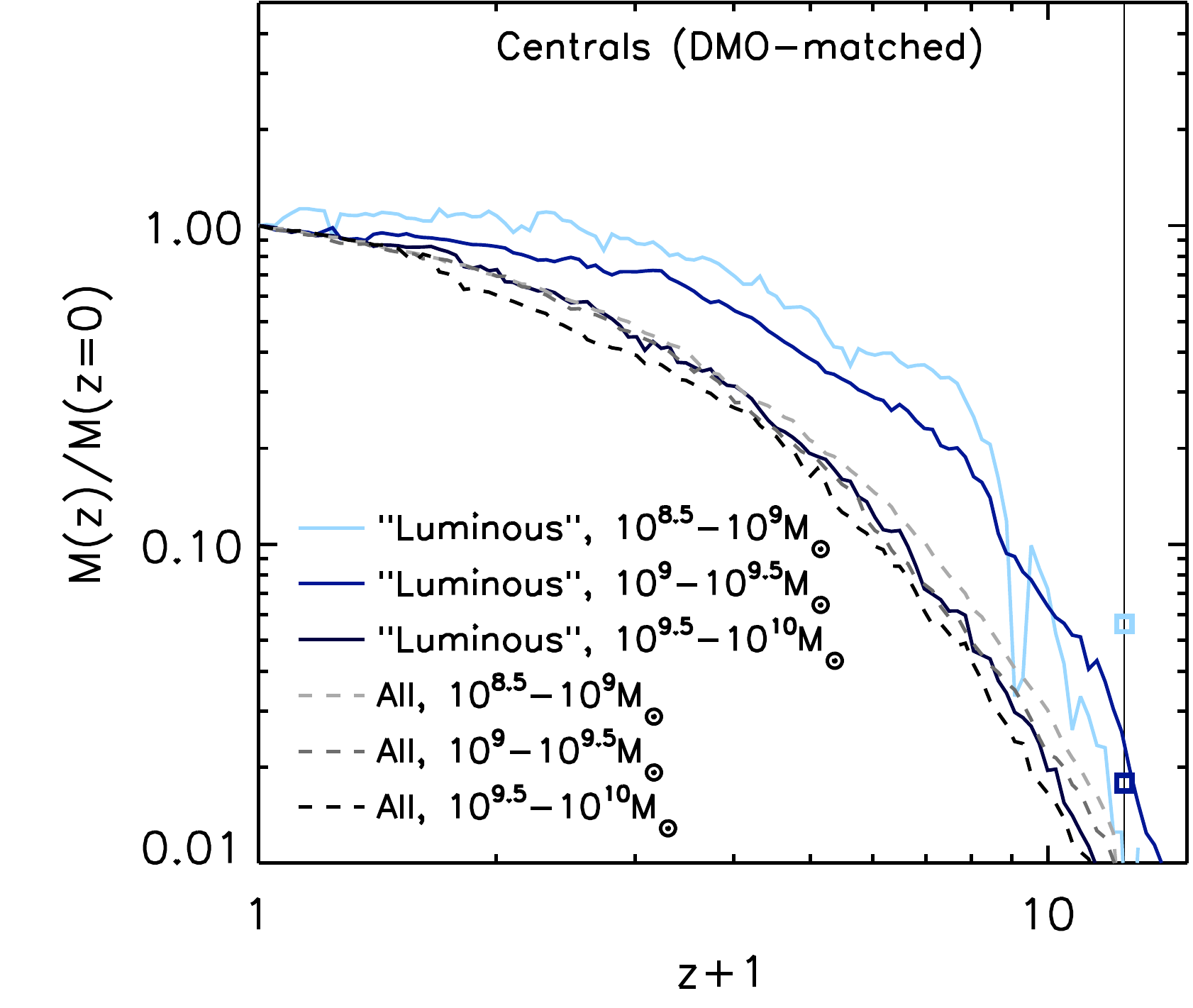}
\includegraphics*[trim = 55mm 135mm 25mm 0mm, clip, height = 56.9mm]{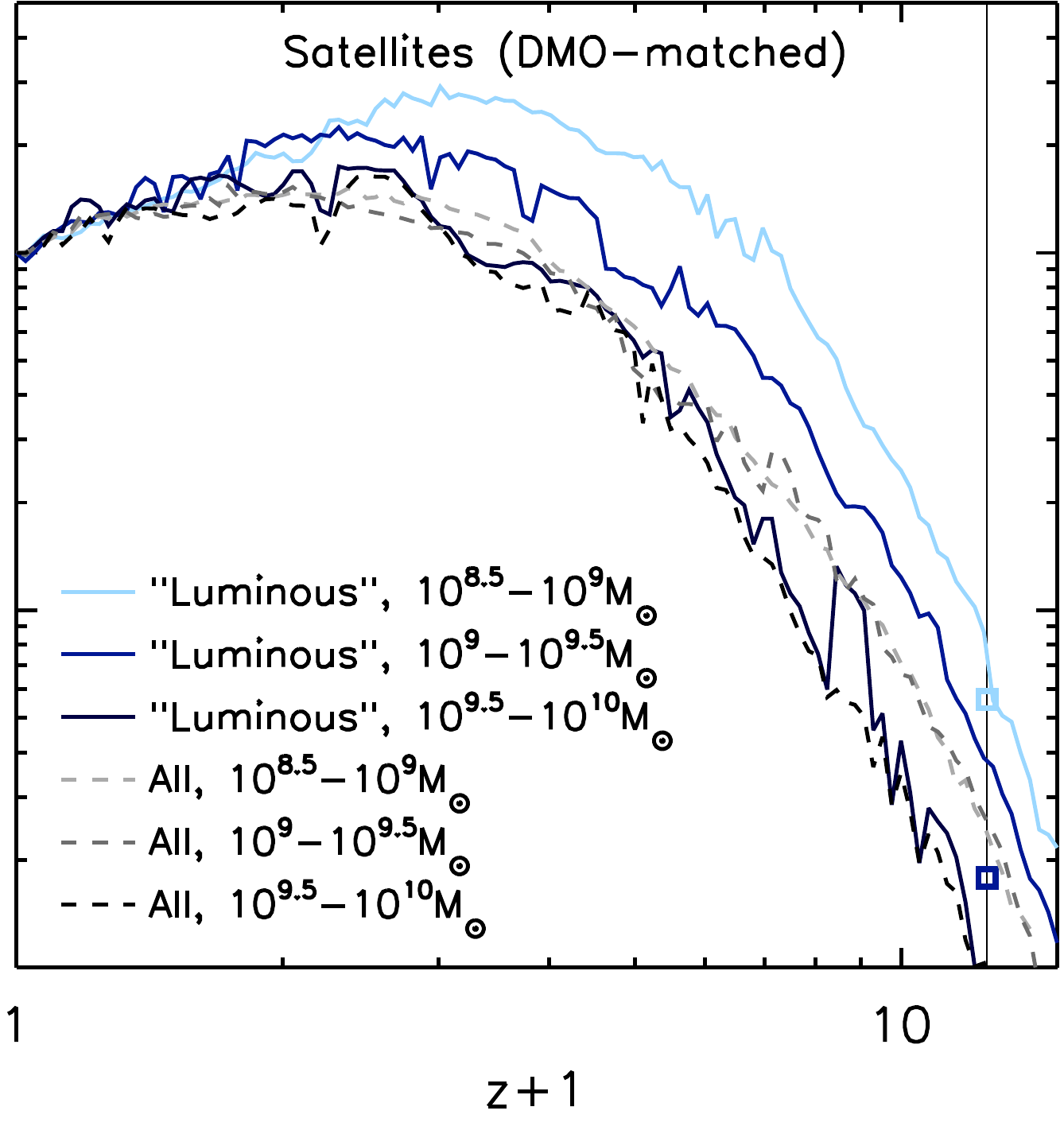} 
\includegraphics*[trim = 55mm 135mm 25mm 0mm, clip,height = 56.9mm]{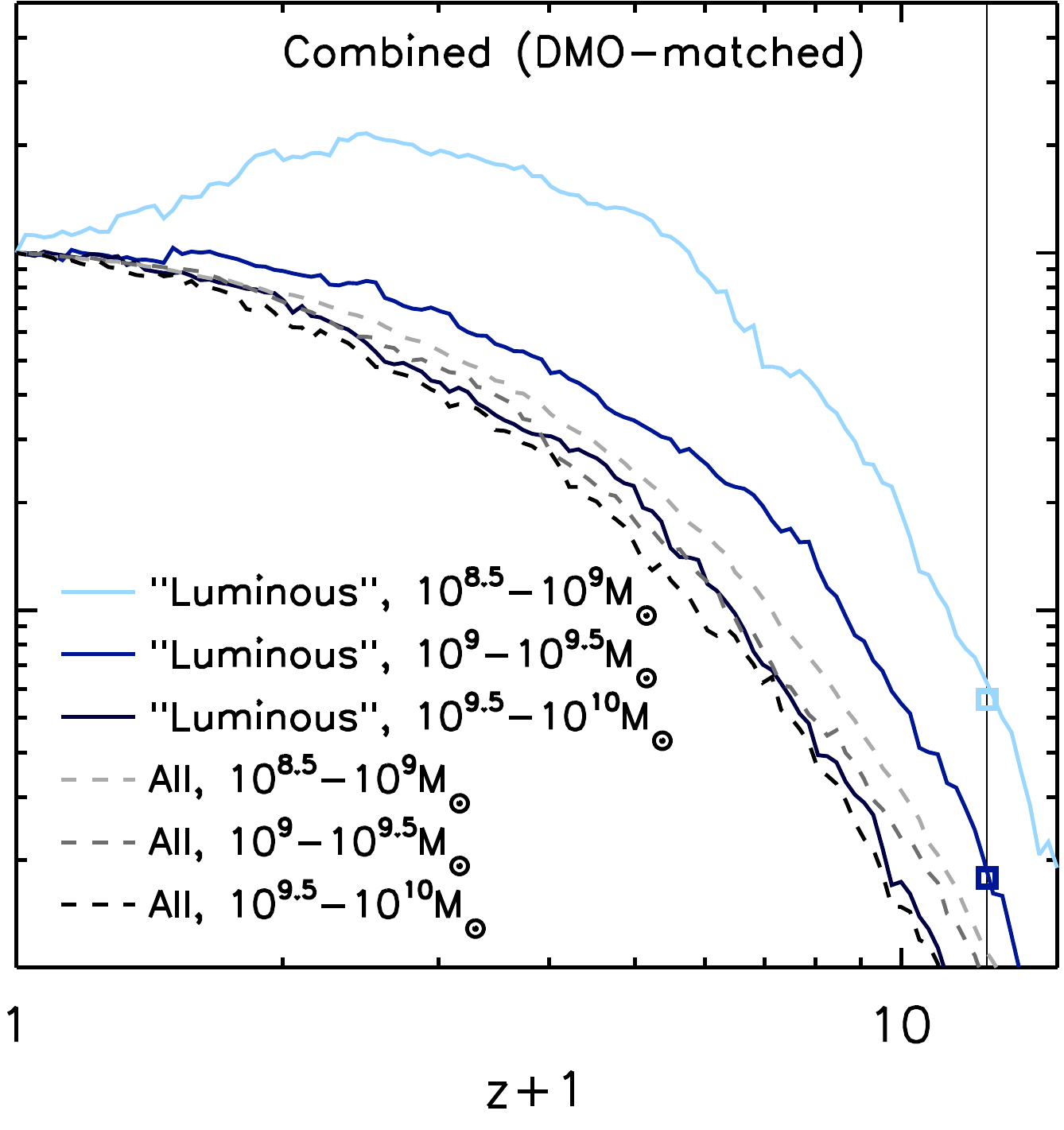} 
\end{center}
\vspace{-.12in}
\caption{Mass assembly history for present-day field halos (left),
  present-day satellites (centre), and the combination of both
  (right). Halos are grouped according to mass at $z=0$, as measured
  in the hydrodynamic simulation (top row), and as measured in the DMO
  simulation (bottom row). The vertical lines indicate $z=11.5$, the
  time of hydrogen reionization, and the squares indicate mass ratios
  that corresponds to the same halo mass ($10^{7.5}\Ms$) at this time
  for each final mass bin. As indicated by the dashed lines, halos of
  lower final mass generally formed earlier, but this trend is greatly
  amplified for luminous halos. With decreasing final mass, luminous
  halos become increasingly biased towards earlier formation
  times. Tidal stripping typically results in mass loss for
  satellites, whereas field halos tend to reach their peak mas at
  $z=0$. However, note that luminous field halos with masses below
  $10^9\Ms$ today are also likely to have been tidally stripped in the
  past, leading to an evolution more similar to that of satellites.}
  \label{fig:assembly-history} 
\end{figure*}

In Fig.~\ref{fig:luminous-fraction} we compare the fraction of halos
containing stars at $z=0$, for simulations of the same volume with and
without reionization and at different resolutions, as defined in
Table~\ref{table:params}. It can be seen that significantly more halos
are luminous in the simulations without reionization. It should be
noted that our simulations are not sufficient to simulate a Universe
without reionization faithfully: the level of fragmentation of the IGM
is limited by resolution, and the total number of galaxies formed in
this (unphysical) scenario is not converged and increases with
increasing resolution. By contrast, the results with reionization are
well converged at L2, suggesting that in our simulations, reionization
sets a limit for star formation in low mass halos that is above the
resolution limit of our simulations.

Fig.~\ref{fig:evolution-luminousfraction} shows the evolution of the
luminous fraction of halos as a function of mass in a simulation that
includes reionization, from $z=11.5$ to $z=0$. At all redshifts, the
fraction of luminous halos at $10^8\Ms$ is less than $10\%$, and
almost no halos are luminous below $10^{7.5}\Ms$. The mass-dependence
of the luminous fraction is strongest at $z=11.5$, where all halos
more massive than $3\times10^8\Ms$ contain stars. As halos typically
grow in mass over time, the mass-dependence of the luminous fraction
becomes more gradual towards lower redshifts. While the minimum halo
mass of luminous halos remains almost unchanged, the mass at which
most halos are luminous increases continuously.

Of course, supernovae and AGN feedback can also heat up the gas and
regulate star formation in larger halos. However, reionization is
clearly a key factor for determining the total number of galaxies that
form and can also strengthen the impact of supernovae on the star
formation rate \citep{Pawlik-2009}. In terms of the halo properties
required for star formation in the presence of reionization, our
results are consistent with those obtained by the previous studies
reviewed in Section~\ref{sec:introduction}. At a mass of $\sim3\times
10^9\Ms$, or $\vm\sim25$~kms$^{-1}$, half of all halos contain
galaxies at $z=0$.

\section{The timing of galaxy formation} \label{sec:timing} Since
reionization takes place at a time when halos have only a small
fraction of their present mass, the probability for star formation
within a halo is expected to depend strongly on its individual
assembly history. Consequently, the properties that separate the halos
that host galaxies from those that remain dark should be more closely
related to their progenitors at high redshifts.

Halos that have assembled more mass at the time of reionization will
be more resilient to photo-evaporation, and will subsequently be able
to cool gas more efficiently, resulting in a greater chance for star
formation. As a result, for a fixed mass today, halos that formed
earlier are more likely to contain galaxies, so that halos that host
galaxies are biased towards earlier formation times.

\subsection{Assembly Histories}\label{sec:assembly}

In Fig.~\ref{fig:assembly-history} we compare the average mass
assembly histories of luminous and dark halos in three different final
halo mass ranges. We distinguish between presently isolated halos and
satellites, and compare the results measured directly in the
hydrodynamic simulation to those measured by comparing the matched
counterparts to the luminous and non-luminous halos from the dark
matter only (DMO) simulation.

As expected, we find that halos which form stars assemble their mass
significantly earlier compared to non-luminous halos of the same mass
today.

We reproduce the well-known result of hierarchical structure formation
that halos of lower mass typically form earlier than halos of larger
mass, independently of whether or not they contain stars. Since most
halos above $10^{9.5}\Ms$ are luminous, we find only a slight
difference between the assembly histories of dark and luminous halos
above this mass. However, as the fraction of luminous halos decreases
in the mass ranges $10^{9}-10^{9.5} \Ms$ and $10^{8.5}-10^{9} \Ms$,
the remaining luminous halos are increasingly biased towards earlier
mass assembly. This result can be readily understood from the fact
that for gas cooling and star formation to take place in the presence
of reionization, a sufficiently high mass needs to have been assembled
at an early time.

It is worth pointing out that while low mass halos have to assemble
their mass early in order to survive reionization, star formation
itself begins later for most halos. Of all halos that contain galaxies
at $z=0$, only $\sim 8\%$ have started forming stars before $z=11.5$,
with a median redshift of $z=6$ for the first stars to form in each galaxy.

As shown in the middle column of Fig.~\ref{fig:assembly-history},
luminous as well as dark satellites of all masses have typically
experienced some degree of tidal stripping and mass loss, leading to a
decrease in average mass towards $z=0$. By comparison, the mass of
present-day field halos is typically maximal at $z=0$. However, the
least massive luminous field halos in the hydrodynamic simulation
achieved their peak mass at $z>0$, and have since lost mass as a
result of tidal interactions. Note that the combination of strong
stripping and ejection means that a pair of matched halos is unlikely
to evolve in the same way across in the hydrodynamic and DMO
simulations, and that low mass centrals in the DMO simulation are more
likely to be matched to slightly more massive centrals in the
hydrodynamic simulation than to be on such a rare orbit. However,
while such objects are rare amongst all halos, star formation in halos
with peak mass of less than $10^9\Ms$ is also rare, so the probability
of an isolated galaxy in such a low mass halo having had such an
exceptional history is strongly increased.

In this context, it is also worth noting that \cite{Teyssier-2012}
have computed the probability that halos of isolated dwarf galaxies
within the Local Group are ``escaped'' satellites, based on orbits of
halos measured in the Via Lactea II dark matter only simulation. They
found that $\sim13\%$ of halos within 1.5 Mpc have passed through the
Milky Way's virial radius. Our results suggest that for actual field
dwarf galaxies in very low mass halos, the probability of past tidal
interactions is significantly higher.

\begin{figure}
  \begin{center}
    \hspace{-.2in}\includegraphics*[trim = 20mm 125mm 20mm 25mm, clip,  width =  .5\textwidth]{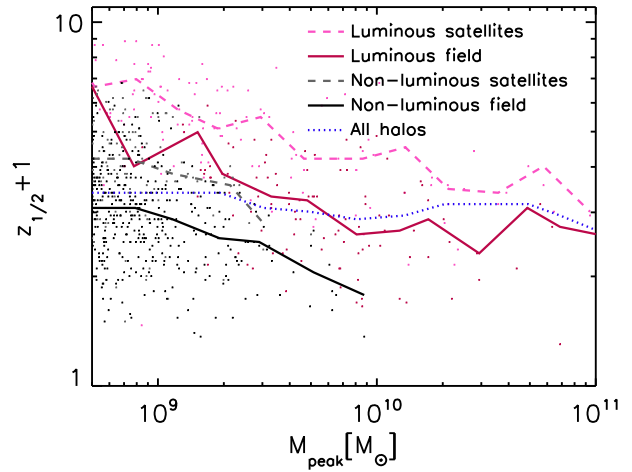}
 \end{center}
\vspace{-.12in}
\caption{Redshift when a halo reached half of its peak mass as a
  function of peak mass. Luminous halos which were able to cool stars
  in the presence of reionization formed significantly earlier than
  non-luminous ones of the same peak mass. As the required evolution
  bias increases for lower mass halos, both luminous and dark halos
  show a strong mass evolution of the formation redshift, whereas the
  total halo population does not.}
  \label{fig:formation-time-bias}
\vspace{-.12in}
\end{figure}

\subsection{Formation Redshifts}\label{sec:formation-redshift}

The difference in assembly history between luminous and non-luminous
halos can also be expressed as a bias in formation redshift,
$z_{1/2}$, defined as the redshift at which a halo's most massive
progenitor first reaches $1/2$ of its peak mass. In
Fig.~\ref{fig:formation-time-bias}, we plot the formation redshifts of
luminous and non-luminous halos as a function of their peak mass.

While the total population of halos shows only a weak dependence of
formation redshift on mass \citep[in agreement with][]{Fakhouri-2010},
the sub-populations of luminous and non-luminous halos {\it both} show
a strong mass-dependence, with lower mass halos forming earlier. This
result can be understood because at any mass, luminous halos form
earlier than dark halos: halos of low mass have to form earlier to be
luminous, and halos of high mass have to form later to be dark. The
increase in the fraction of luminous halos with mass then offsets the
negative correlation of mass with formation redshift in each subset to
give a total population in which mass and formation time are largely
uncorrelated.

The average formation redshift of present-day satellites is higher
than that of present-day field halos. Satellite halos typically reach
their peak mass at $z_{\rm{infall}}$ before $z=0$, while the masses of
field halos typically continue to increase to $z=0$. The exception are
present-day field halos that experienced past tidal interactions,
which, like satellites, are below their peak mass today. As noted in
Section~\ref{sec:assembly}, among field halos with peak masses below
$10^9\Ms$ that are luminous the probability of such a history is
increased significantly.

In summary, we find that the progenitors of present-day dwarf galaxies
do not have the assembly histories typical of dark matter halos of
their mass or $\vm$. While dark halos of peak mass $10^9\Ms$ form at
$z_{1/2}\sim2$, halos of the same mass that host galaxies have a
formation redshift of $z_{1/2}\sim 3-4$.

\begin{figure*}
  \begin{center}
\vspace{-.1in}
\hspace{-.05in}\includegraphics*[trim = 65mm 25mm 60mm 30mm, clip,  height = .44\textwidth]{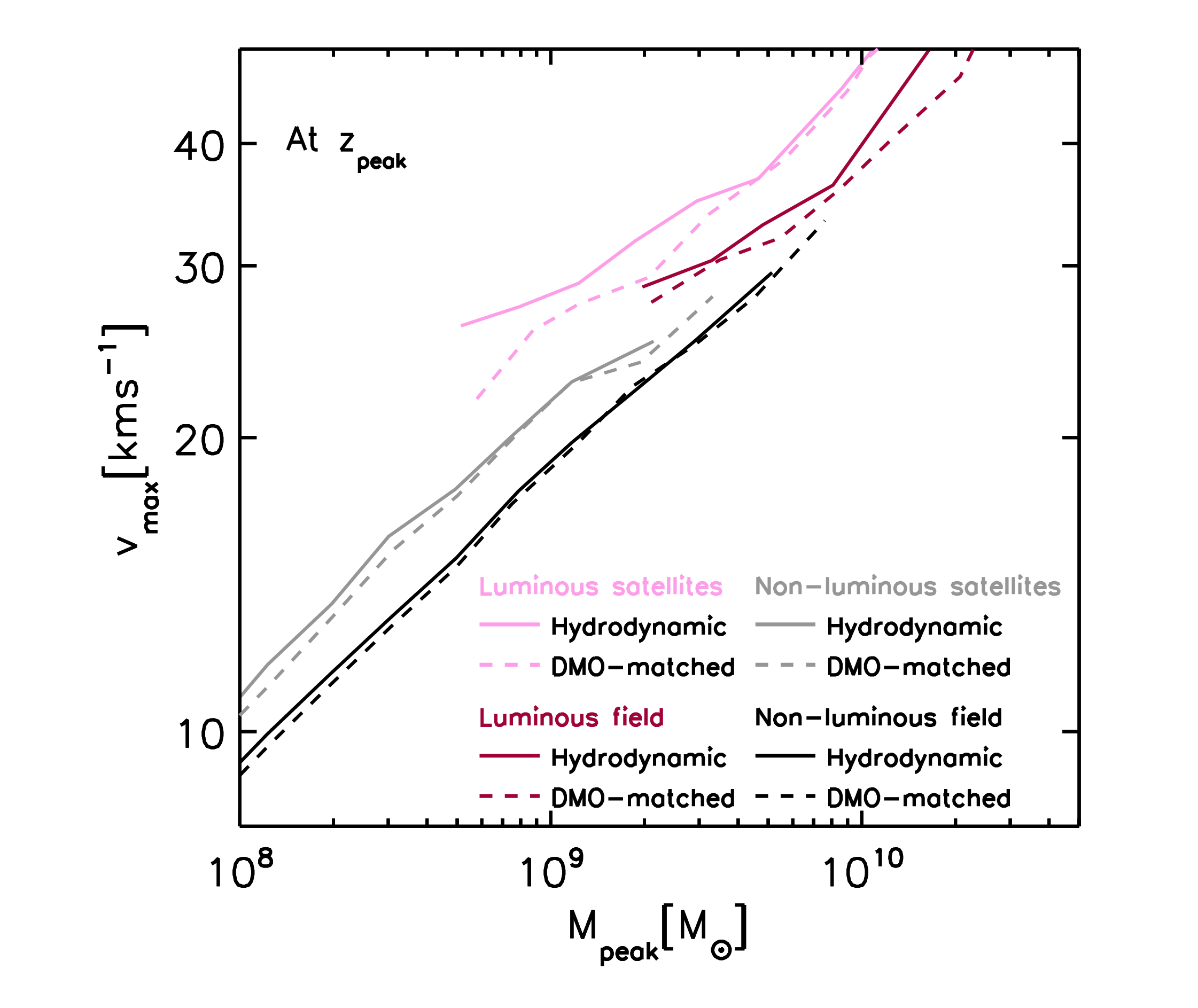}
\hspace{-.05in}\includegraphics*[trim = 110mm 25mm 60mm 30mm, clip,  height = .44\textwidth]{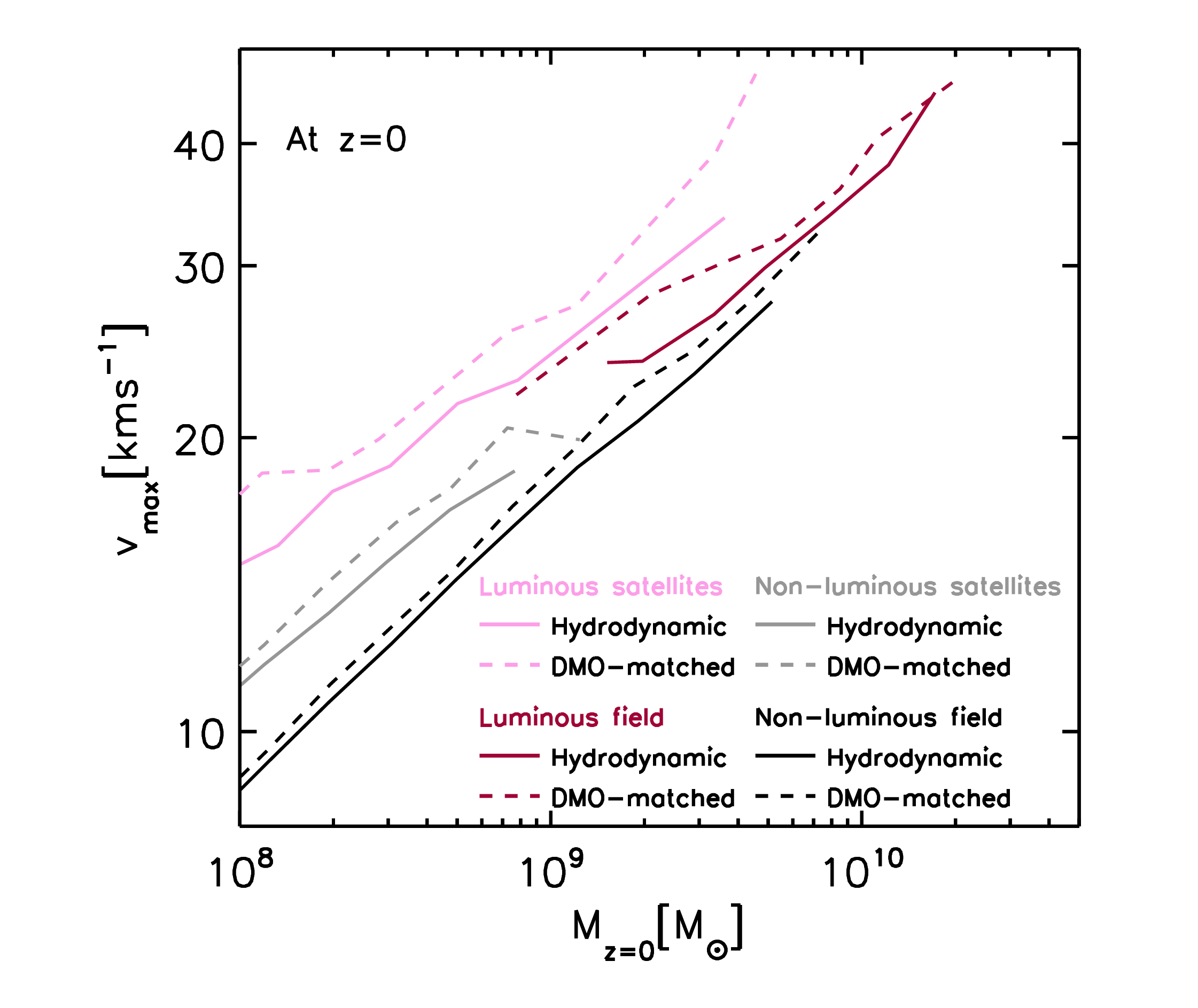}
 \end{center}
\vspace{-.12in}
\caption{The relation between halo mass and $\vm$ at the time of peak
  mass (left panel) and at $z=0$ (right panel) from the simulation at
  resolution L1 including reionization. Since stars form
  preferentially in halos of high concentration, luminous halos are on
  average more concentrated than dark halos. The effect of galaxy
  formation itself on the halo is manifest in the difference between
  the relations measured in the hydrodynamic simulation and those
  measured from the matched DMO simulation. At peak mass, luminous
  halos in the hydrodynamic simulation have higher $\vm$/mass ratios
  than their DMO counterparts, indicating that gas cooling leads to an
  increase in concentration. At $z=0$ halos of all type are typically
  less concentrated in the hydrodynamic simulation than in the DMO
  simulation.}
  \label{fig:mass-vmax}
\vspace{-.05in}
\end{figure*}

\subsection{Velocity~--~mass relation}\label{sec:velocity-mass}
Halos that formed earlier have higher concentration, and therefore
higher maximum circular velocity, $\vm$, for a given mass. In
addition, more concentrated halos can cool gas more efficiently,
limiting the photo-evaporating effect of reionization. Since both
early mass growth and the resistance to photo-evaporation enhance the
probability for star formation, we expect low-mass galaxies to be
hosted preferentially by halos of higher $\vm$~--~mass ratios.

In Fig.~\ref{fig:mass-vmax} we show the relation between $\vm$ and
halo mass, either evaluated at $z=0$ or at $z_{peak}$, the time of
peak mass. While the total population follows the $\vm$~--~mass
relations expected for $\Lambda$CDM \citep[e.g.][not
shown]{Klypin-2011}, we find that below $10^{9.5}\Ms$, luminous halos
have significantly higher $\vm$~--~mass ratios than non-luminous
ones. As the fraction of luminous halos in our hydrodynamic simulation
decreases with decreasing halo mass, the bias of the remaining
luminous halos increases. Similarly, at high masses where most halos
are luminous, the remaining dark halos are increasingly biased towards
lower $\vm$~--~mass ratios.

We find similar trends when we consider the halos of the DMO
simulation that are matched to luminous and non-luminous halos. Like
the luminous halos themselves, their counterparts in the DMO
simulation have much higher $\vm$~--~mass ratios than the DMO
counterparts of the non-luminous halos. This confirms that the
increased $\vm$ of luminous halos in the hydrodynamic simulation is
largely explained by the fact that more concentrated halos are
intrinsically more likely to form stars.

We can examine the additional effect of baryons by directly comparing
the $\vm$~--~mass relation between halos in the hydrodynamic
simulation and their respective DMO counterparts. From the left panel
of Fig.~\ref{fig:mass-vmax}, it can be seen that at peak mass, the
non-luminous halos and their DMO counterparts follow very similar
relations, while the luminous halos in the hydrodynamic simulation are
more concentrated than their DMO counterparts. This indicates that
processes like gas cooling, which are stronger in the luminous than in
the the non-luminous halos, can also lead to an increase in the
$\vm$~--~mass ratios, an effect obviously not present in the DMO
simulation.

When the two simulations are compared at $z=0$, as is the case in the
right panel of Fig.~\ref{fig:mass-vmax}, luminous and non-luminous
halos alike have lower average $\vm$~--~mass ratios than their
respective counterparts in the DMO simulation. However, this
difference is small compared to the offset between luminous and dark
halos or the corresponding matched objects.

It can also be seen that satellites follow a different $\vm$~--~mass
relation compared to field halos, with a lower mass for a given
$\vm$. This well-known result can be attributed to the fact that tidal
stripping first removes material from the outside of an infalling
halo, beyond $\rmax$, with a greater impact on halo mass than on
$\vm$. However, we also find that the difference between satellite and
field halos is strongly amplified among luminous halos because, as we
will discuss in Section~\ref{sec:satellites}, luminous and dark
satellites follow significantly different orbits.

In summary, we find that luminous low-mass halos have much higher
$\vm$~--~mass ratios than average halos. This difference increases as
the fraction of luminous halos decreases towards lower masses, and is
higher for satellites than for field halos. Since we find similar
trends between the respective matched halos in DMO simulation, we
attribute them mostly to the increased likelihood for star formation
in more concentrated halos, rather than to an increase of
concentration due to cooling and star formation.

\begin{figure*}
  \begin{center}
\vspace{-.05in}
    \hspace{-.65in}\includegraphics*[trim = 5mm 155mm 27mm 0mm, clip, height=47.45mm]{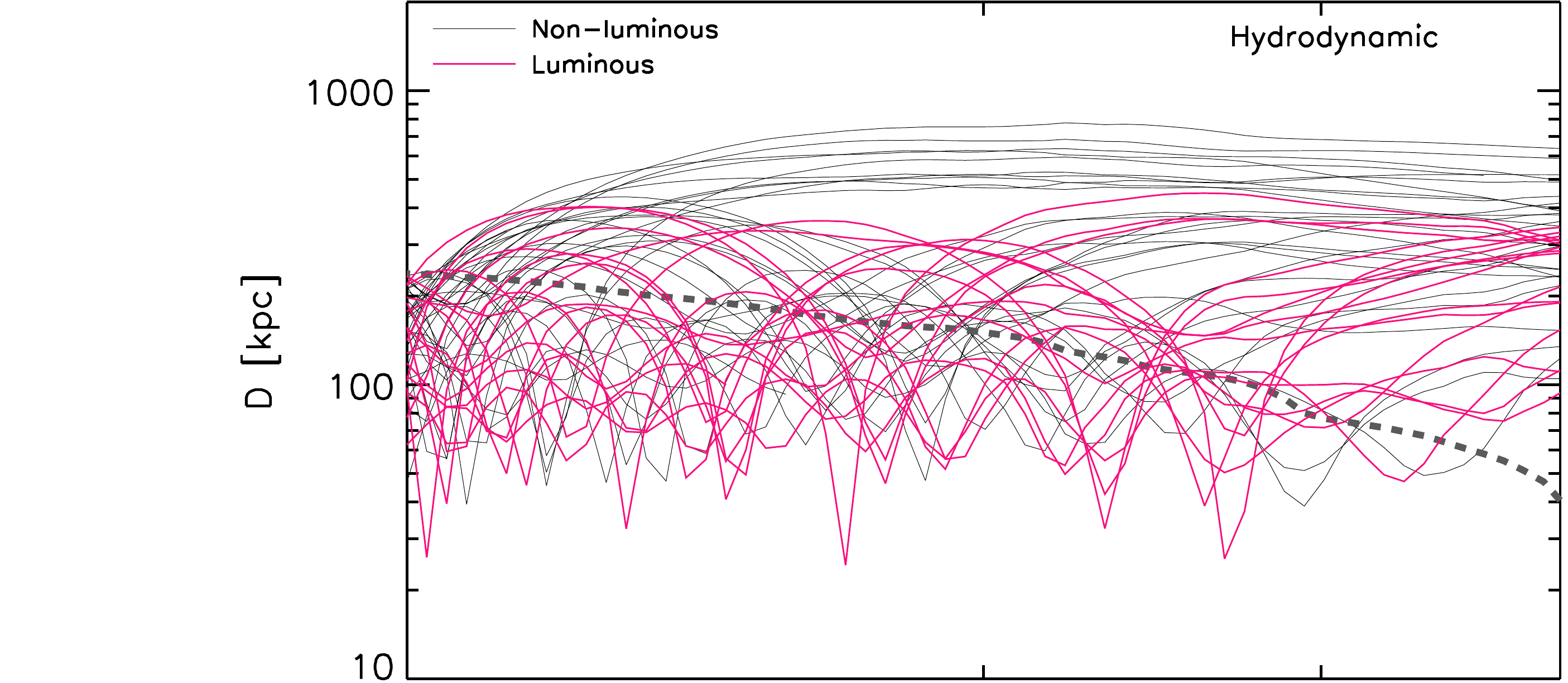}
    \includegraphics*[trim = 77mm 155mm 27mm 0mm, clip, height =  47.45mm]{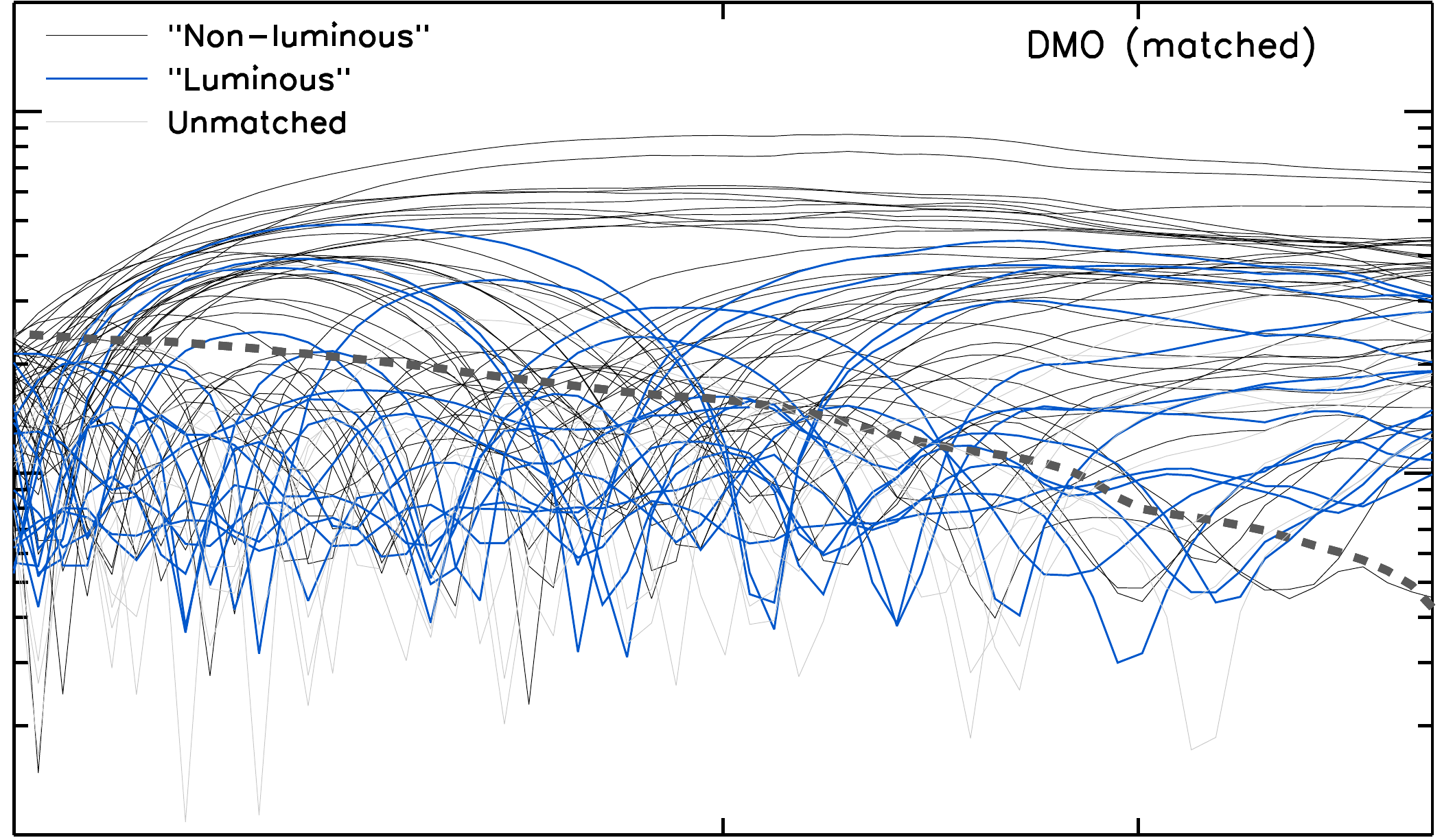} \\
    \hspace{-.65in}\includegraphics*[trim = 5mm 130mm 27mm 0mm, clip, height = 57mm]{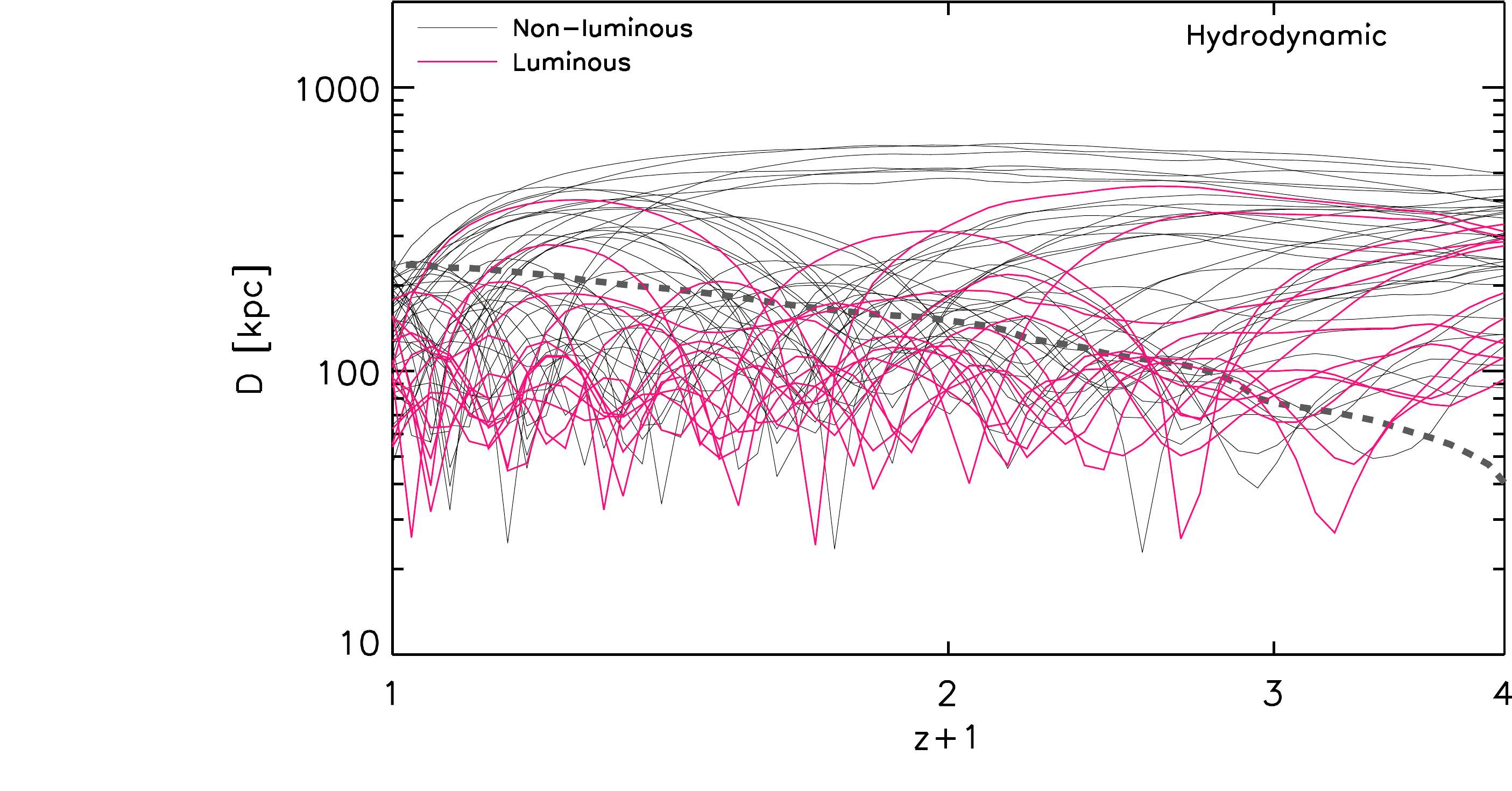}
    \includegraphics*[trim = 77mm 130mm 27mm 0mm, clip,  height = 57mm]{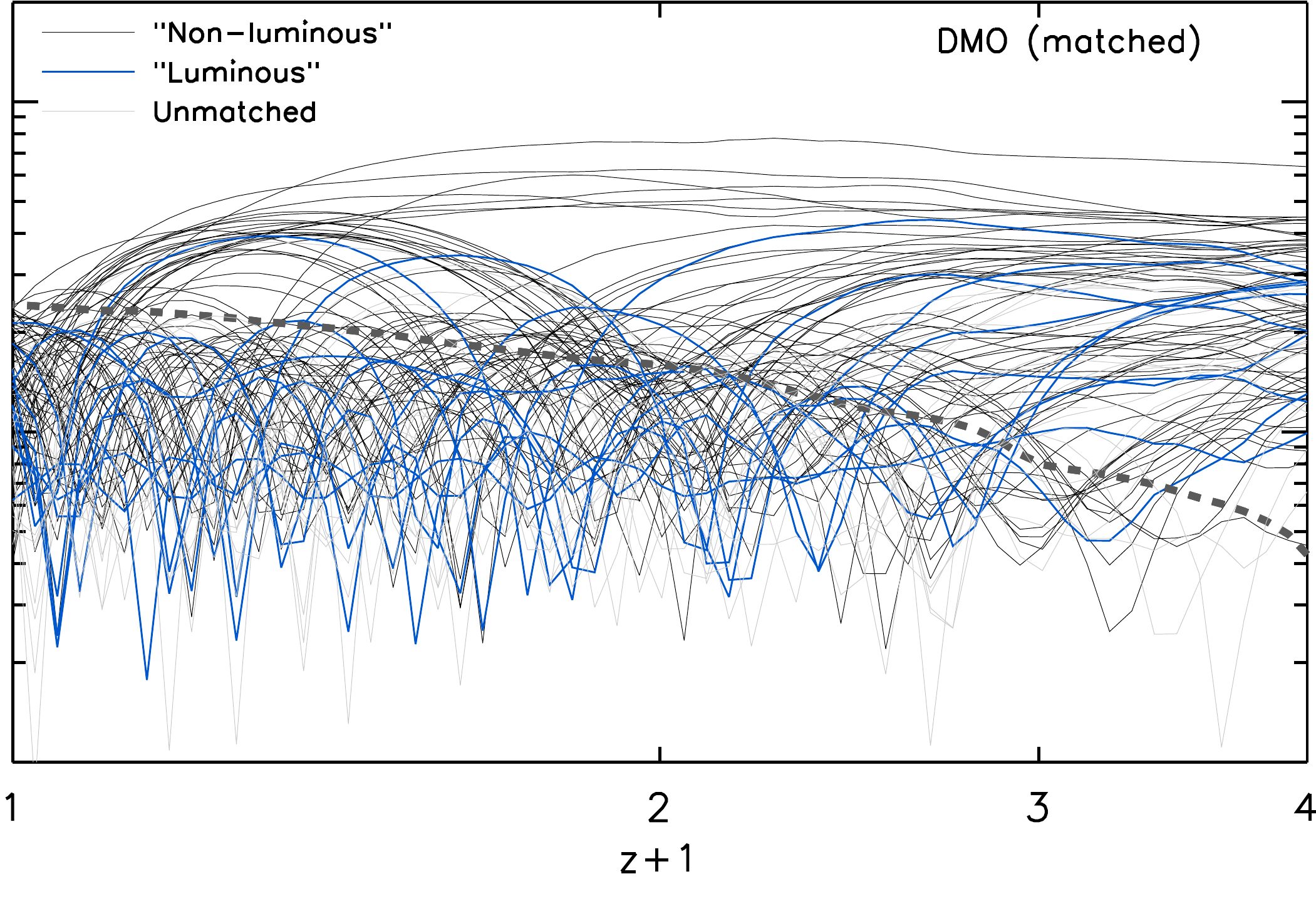}
    
 \end{center}

\vspace{-.17in}

\caption{Evolution of galactocentric distance for dark and luminous
  satellites of the simulated ``Milky Way'', as identified directly in
  the hydrodynamic simulation (left column), or as matched from the
  DMO simulation (right column) at resolution L1. Satellite halos are
  selected by present mass in the range $10^8-10^9\Ms$ (top row) or
  present $\vm$ in the range $12-20$~kms$^{-1}$ (bottom row). In each
  panel, the grey dashed line shows the evolution of the host halo's
  virial radius, $\mathrm{r_{200}}$, in both simulations. Within this
  range of mass or $\vm$, the orbits of luminous and dark halos differ
  significantly. Also note the lower number of objects in the
  hydrodynamic simulation compared to the DMO simulation, due to the
  reduction of $\vm$ and mass of individual halos by baryonic
  effects.}
  \label{fig:satellite-orbits}
\vspace{-.03in}
\end{figure*}

\section{Late Time Evolution}\label{sec:bias}
In addition to the effects of assembly history and the properties of
halos directly related to reionization, the late-time evolution also
influences the probability for halos to host galaxies, and the
expected distribution of luminous and dark halos within the Local
Group.

In the previous section, we showed that the progenitors of luminous
halos formed significantly earlier than those of dark halos. We also
noted clear differences between the evolution of satellites and
centrals. These are not a direct consequence of reionization, but
indicate that differences in the late-time evolution can change the
correspondence between the properties of present-day halos and those
of their progenitors in the early universe.

In this section, we examine the late-time effects in more detail, and
show in particular how tidal stripping and the fact that satellites
stop growing after infall, can introduce a strong environmental
dependence on the relation between observable galaxies and the
underlying halo population.

\subsection{Satellite Evolution} \label{sec:satellites} The evolution
of satellite halos is driven by tidal effects, which further separate
their present-day properties from those of their progenitors. Among
satellite halos of the same mass or $\vm$ today, those that
experienced greater mass loss had more massive progenitors prior to
infall, and therefore a higher probability for star formation. This
introduces additional biases between dark and luminous satellites that
depend on their infall times and orbital parameters.

In Fig.~\ref{fig:satellite-orbits} we compare the orbits of luminous
and dark low-mass satellites of one of our simulated Milky-Way like
halos. We identify satellites as self-bound halos within the virial
radius of the host at $z=0$, $\mathrm{r_{200}}$, defined as the radius within
which the mean density is 200 $\times$ the {\it critical} density, and
define the infall redshift, $z_{\rm infall}$, as the redshift when a
satellite first crosses $\mathrm{r_{200}}$. Selecting satellites either by
their present mass or present $\vm$, it can be seen that the orbits of
luminous halos in the mass range of $10^8-10^9~\Ms$ or the $\vm$ range
of $12-20$~kms$^{-1}$ differ significantly from those of dark halos of
the same present mass: they fall in earlier and come closer to the
centre of the host.

We note that the same effect can be seen between the halos of the DMO
simulation which are matched to luminous and dark halos in the
hydrodynamic simulation and shown in the right column of
Fig.~\ref{fig:satellite-orbits}. The matching of low mass satellites
is imperfect because satellites can evolve differently in the two
simulations. Nevertheless, it underlines the fact that the different
orbits of luminous and non-luminous satellites are primarily caused by
a selection effect of the progenitors imposed by reionization, and not
by differences in the baryonic components between the luminous and
dark satellites. However, the overall number of satellites is reduced
in the hydrodynamic simulation due to the loss of baryons, consistent
with the results of \cite{Sawala-abundance}.

\begin{figure}
  \begin{center}
\vspace{-.1in}
 \includegraphics*[trim = 30mm 130mm 10mm 62mm,
  clip, width =
  .48\textwidth]{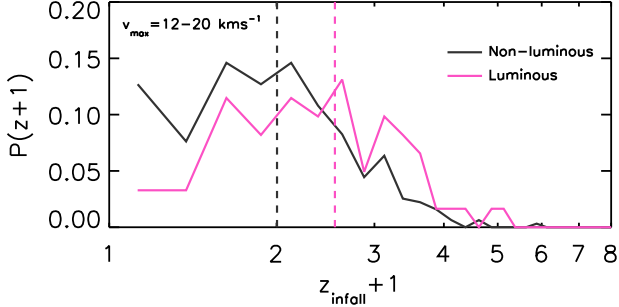}  \\
\vspace{-.2in}
 \end{center}
 \caption{Distributions of infall redshifts of luminous and
   non-luminous satellites in the present-day $\vm$ range of
   12--20~kms$^{-1}$. The dashed lines indicate the median
   values. Luminous halos of these $\vm$ values fell in significantly
   earlier than non-luminous halos.}
  \label{fig:satellite-infall-redshifts}
\end{figure}

\begin{figure}

  \begin{center}
   \hspace{-.15in}\includegraphics*[trim = 25mm 130mm 10mm 62mm,    clip, width =    .48\textwidth]{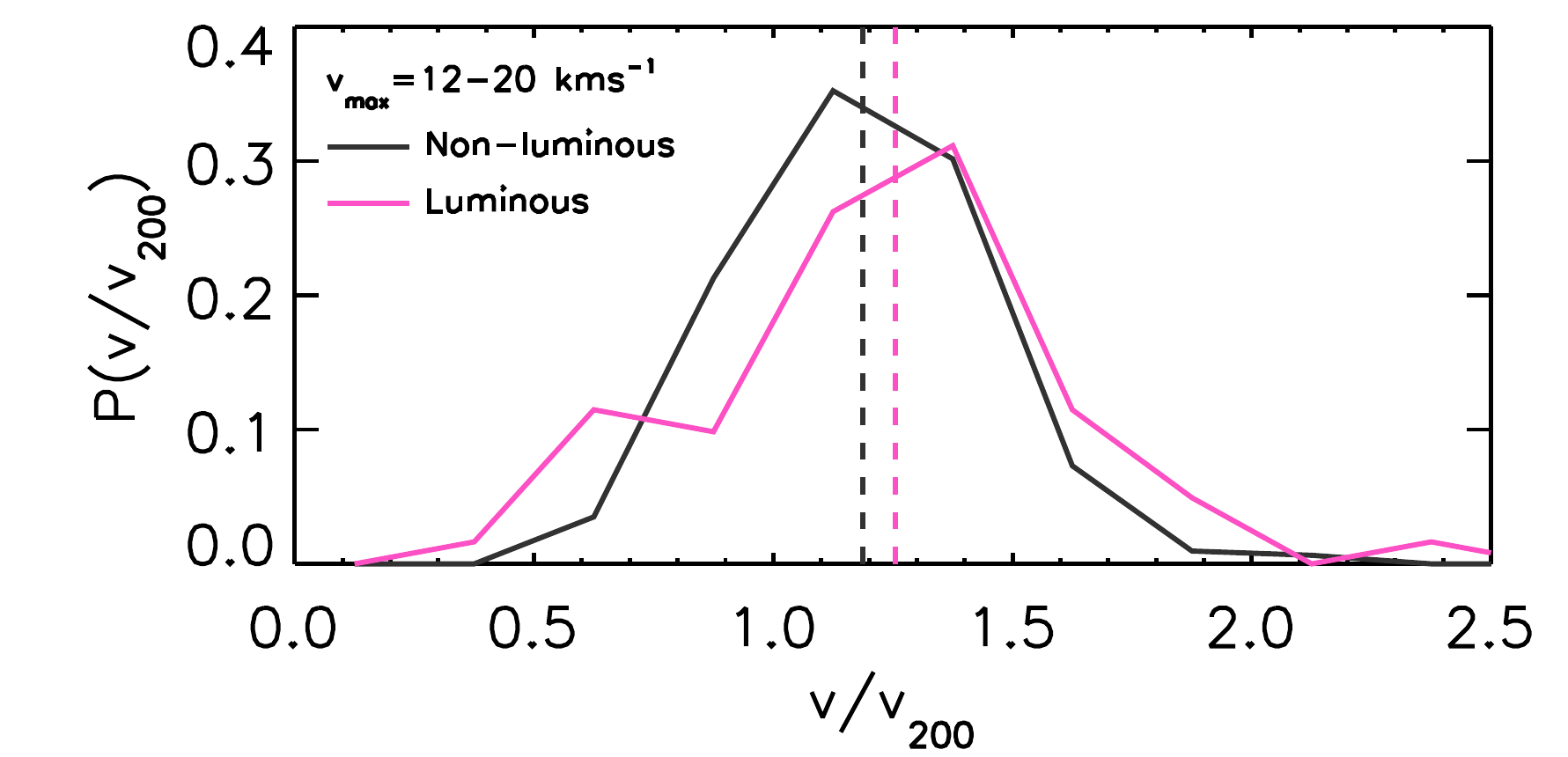} \\
\vspace{-.02in}
    \hspace{-.15in}\includegraphics*[trim = 25mm 130mm 10mm 62mm,    clip, width =    .48\textwidth]{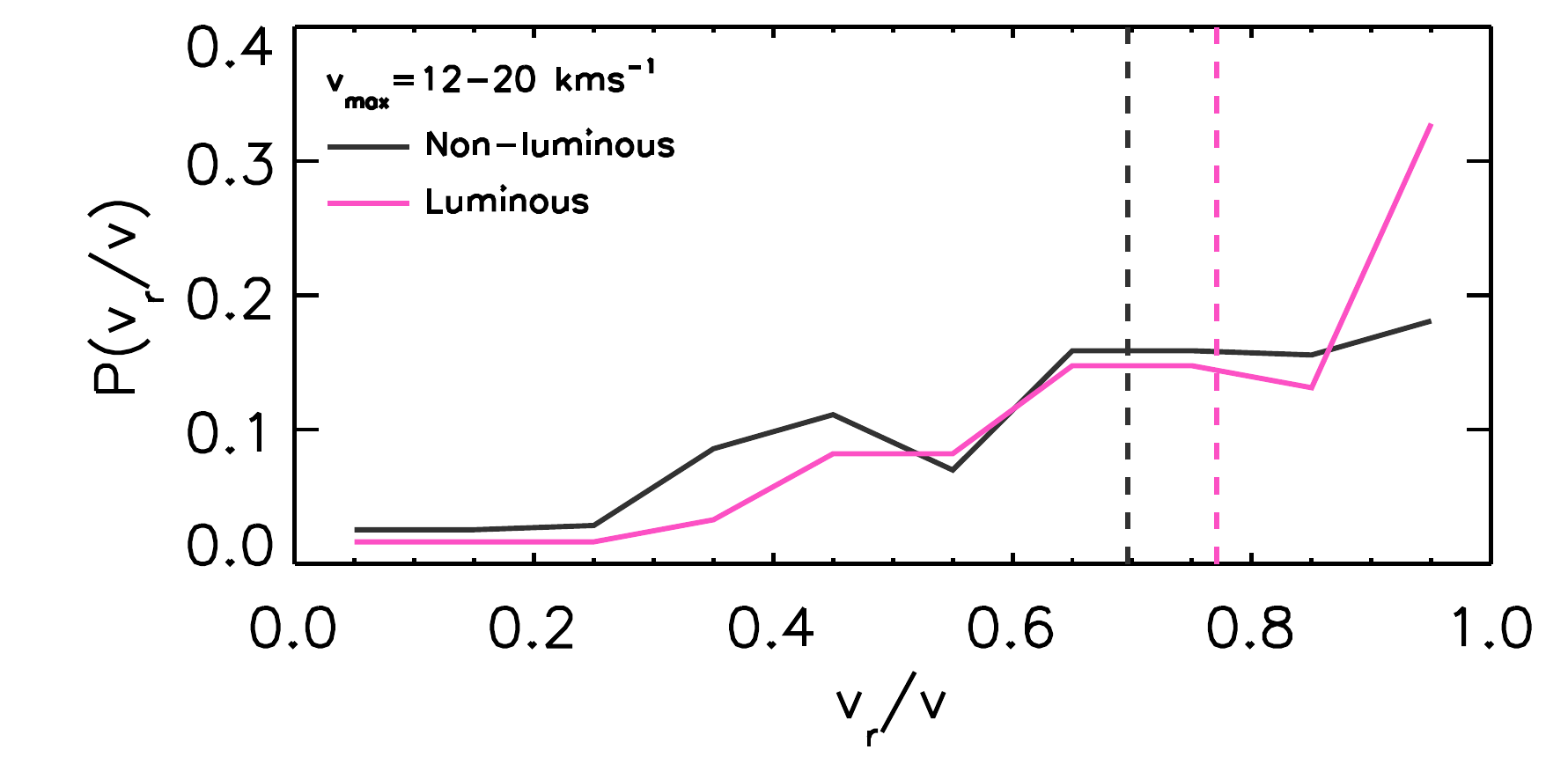}
 \end{center}
\vspace{-.15in}
\caption{Top panel: distributions of infall velocities of satellites
  in the present-day $\vm$ range of 12--20~kms$^{-1}$ divided by
  $\mathrm{v_{200}}$ of the host halo at the time of infall. Bottom
  panel: distributions of the ratio between the radial component and
  the total infall velocity. On both panels, the dashed lines indicate
  the median values. Luminous halos in this $\vm$ range fell in with
  slightly higher velocities and on more radial orbits compared to
  non-luminous halos.}
  \label{fig:satellite-orbital-parameters}
\end{figure}

Fig.~\ref{fig:satellite-infall-redshifts} shows the distribution of
infall redshifts of luminous and non-luminous satellites in the $\vm$
range of $12-20$~kms$^{-1}$, comparable to the values inferred for the
halos that host the Milky Way dwarf spheroidal galaxies. These results
include satellites of all 10 M31 and Milky Way like halos in our five
Local Group simulations at resolution L2. The median infall redshift
of non-luminous satellites in this $\vm$ range is 1.0, while luminous
satellites fell in significantly earlier, with a median infall
redshift of 1.5.

It is worth noting that in addition to {\it mass loss} through tidal
stripping, the infall of a satellite also marks the end of {\it mass
  growth} that a field halo would typically experience. For the same
infall mass, an earlier infall time corresponds to a higher mass in
the early universe, further enhancing the probability of star
formation.

In Fig.~\ref{fig:satellite-orbital-parameters} we compare the same
population of satellites described above in terms of the ratio between
the infall velocity of the satellite and the virial velocity of the
host halo, $\mathrm{v/v_{200}}$, at the time of infall, and in terms of the
ratio between the radial component and the total infall velocity,
$\mathrm{v_r/v}$. Again, we find that luminous satellites in this $\vm$-range
fell in with slightly higher infall velocities and on more radial
orbits, both of which made them more susceptible to tidal stripping.

\begin{figure}
  \begin{center}
 \vspace{-.1in}
    \hspace{-.2in}\includegraphics*[trim = 15mm 125mm 20mm 25mm, clip,   width = .5\textwidth]{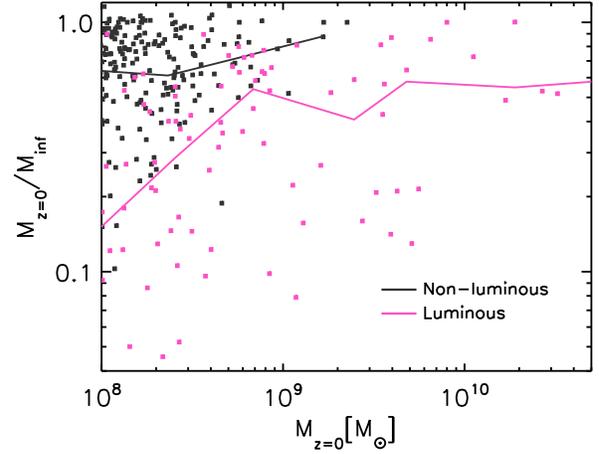} \\
    \hspace{-.2in}\includegraphics*[trim = 15mm 125mm 20mm 25mm, clip,   width = .5\textwidth]{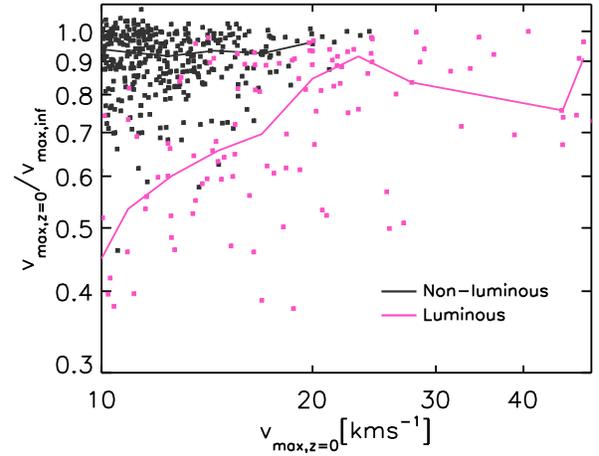} 
 \end{center}
\vspace{-.2in}
\caption{Change in mass relative to mass at $z_{\rm infall}$ as a
  function of present mass (top panel), and change in $\vm$ relative
  to $\vm$ at $z_{\rm infall}$ as a function of present $\vm$ (bottom
  panel), for luminous and non-luminous satellite halos of the parent
  halo shown in Fig.~\ref{fig:satellite-orbits}. Lines indicate the
  median values. As halos with infall masses $< 10^{9.5}\Ms$ or $\vm <
  25$~kms$^{-1}$ are unlikely to host galaxies, luminous halos with
  such lower present values are likely to have experienced a much more
  significant decrease in mass and $\vm$ from their peak values
  compared to non-luminous halos. }
  \label{fig:present-vs-infall}
\end{figure}

The post-infall changes in mass and $\vm$ among luminous and dark
satellites as functions of present mass and $\vm$ are compared in
Fig.~\ref{fig:present-vs-infall}. As expected, luminous satellites of
low present mass are increasingly more likely to have lost mass due to
stripping. Whereas dark satellites have typically lost only $1/3$ of
their mass after infall, independent of present mass, luminous
satellites of $10^9\Ms$ have lost of 1/2 of their mass in the median,
a fraction that increases further for lower mass halos. While dark
satellites have typically reduced their $\vm$ by less than $10\%$,
luminous satellites of 20~kms$^{-1}$ today have typically experienced
a reduction in $\vm$ by $25\%$.

The bias seen in the infall times, orbits, and reduction in mass and
$\vm$ of satellites similar to those that host the observed dwarf
spheroidal galaxies may change previous assumptions about their
evolution: whereas \cite{Penarrubia-2008} have shown using DMO
simulations that the $\vm$ of most dwarf-sized {\it halos} should not
be affected by tidal effects, our simulations show that the opposite
is be true for the subset of halos that actually host the dwarf
spheroidal {\it galaxies}. Those satellite galaxies that today inhabit
halos of $\vm \sim 20$~kms$^{-1}$ typically formed in halos of $\sim
30$~kms$^{-1}$.

\begin{figure*}
  \begin{center}
\vspace{-.2in}
    \hspace{-.2in}\includegraphics*[trim = 0mm 90mm 2mm 60mm, clip,
    height = .64 \textwidth]{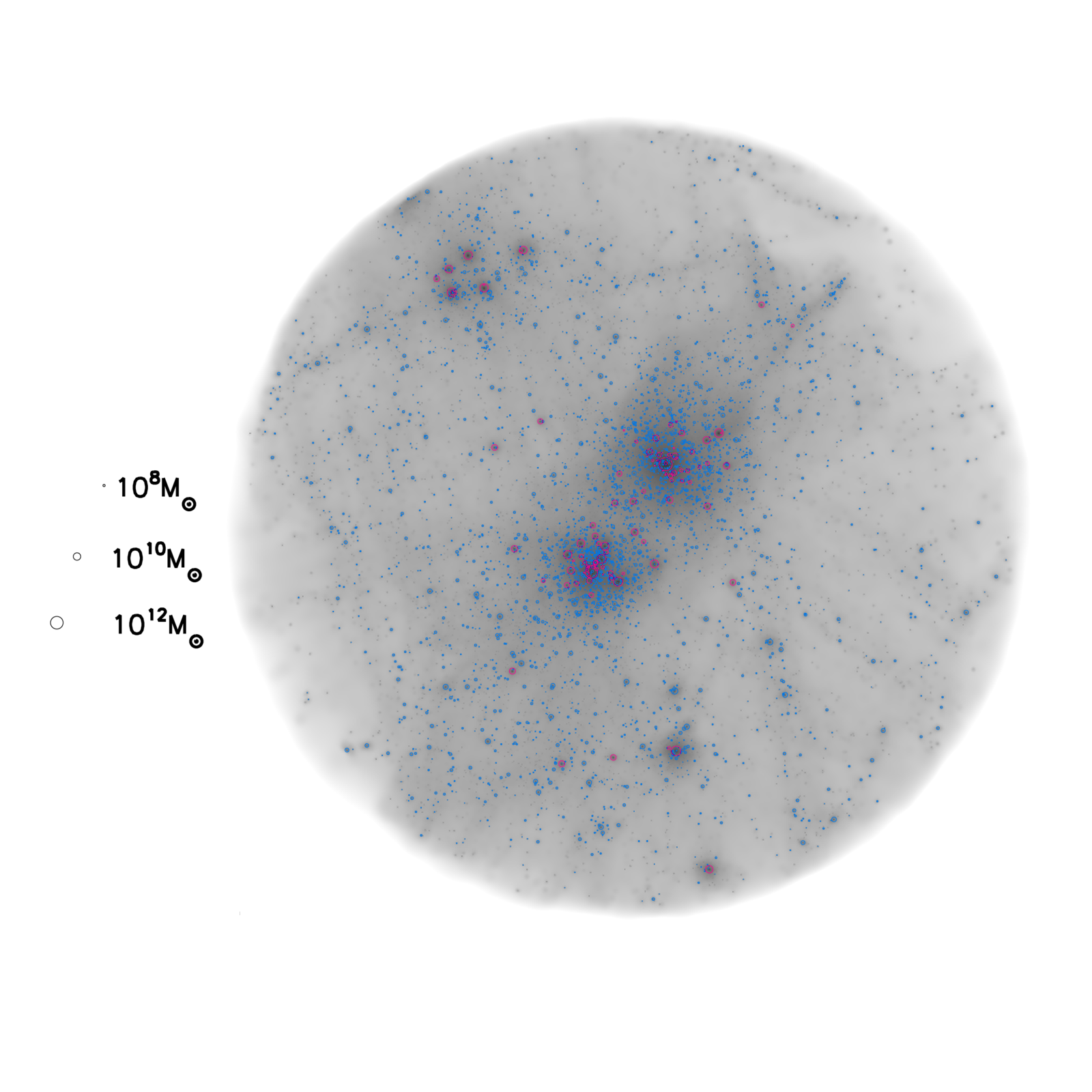} \\
    \hspace{-.2in}\includegraphics*[trim = 0mm 80mm 2mm 60mm, clip,
    height = .64 \textwidth]{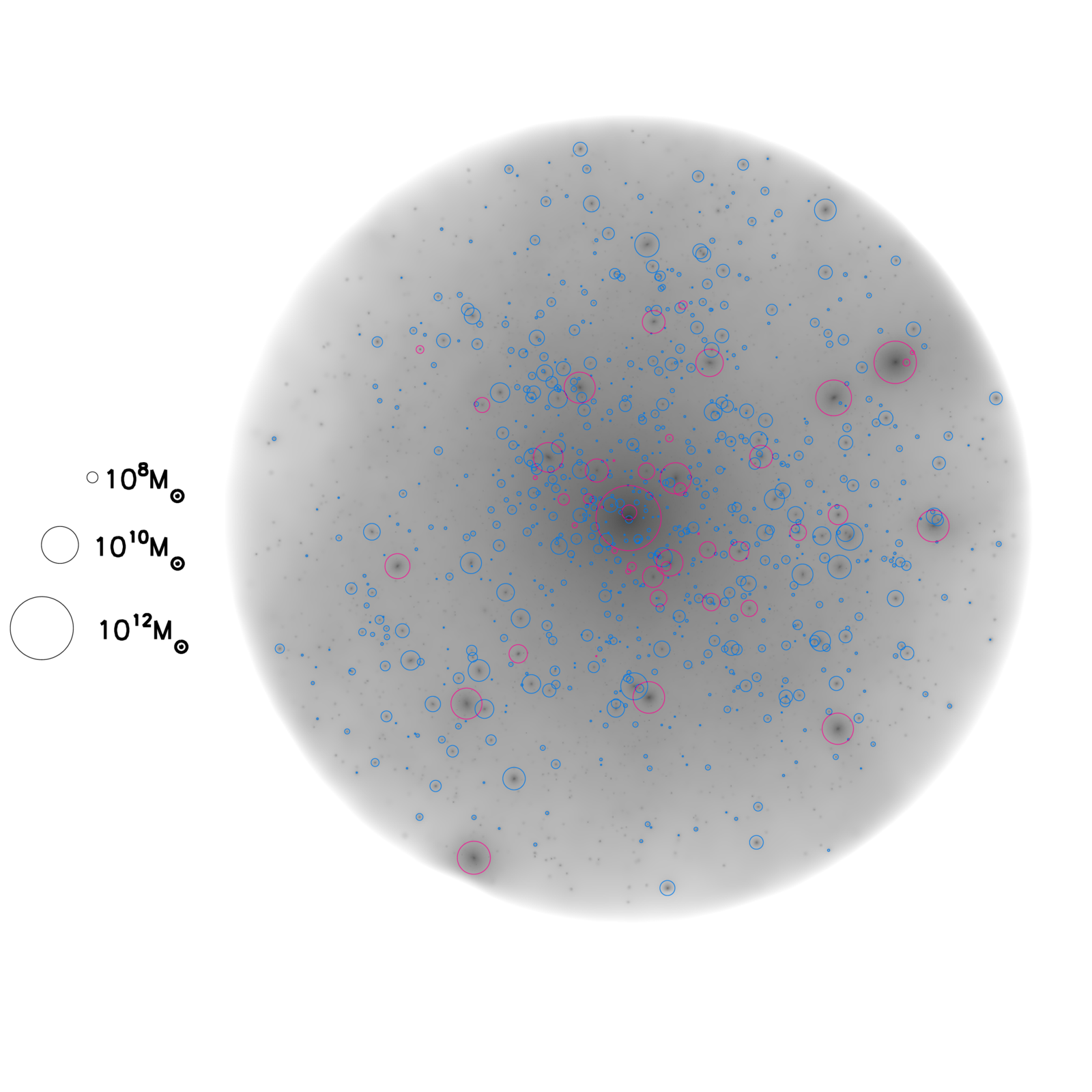}
\vspace{-.2in}

  \end{center}
  \caption{Projected density distribution of dark matter and positions
    of halos at $z=0$ in the same simulated Local Group volume as
    Fig.~\ref{fig:baryon-density}, in a sphere of radius 2~Mpc centred
    on the LG barycentre (top panel), and a sphere of 400 kpc radius
    centred on the simulated Milky-Way (bottom panel).  Blue and red
    circles indicate the positions of dark and luminous halos above
    $10^7\Ms$, respectively, with diameter proportional to the
    logarithm of the mass. It can be seen that luminous halos of low
    mass cluster near the two main halos. Also, while more massive
    halos are more likely to be luminous, there is no sharp mass
    threshold separating luminous and dark halos at $z=0$, since the
    probability of hosting a galaxy is a function of mass, assembly
    history, and environment.}
  \label{fig:matter-density}
\end{figure*}

\begin{figure}
  \begin{center}
\vspace{-.05in}
    \hspace{-.1in}\includegraphics*[trim = 30mm 120mm 15mm 35mm, clip, width = .48\textwidth]{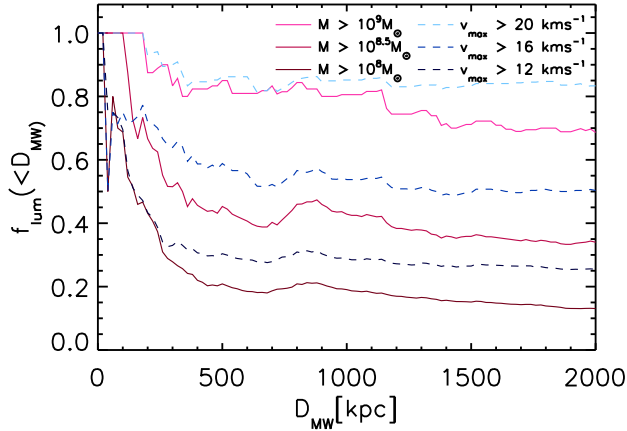}
  \end{center}
\caption{Fraction of halos above a given mass (solid lines) or $\vm$
  (dashed lines), within a given distance from the centre of the
  simulated ``Milky Way''. At fixed mass, halos within $\sim 300$ kpc
  of the Milky Way are much more likely to contain galaxies than more
  isolated halos, and the environmental dependence is stronger among
  halos of lower present mass or $\vm$.}
  \label{fig:dark-fraction-distance}
\end{figure}

\subsection{Dwarf Galaxies and Halos in the Local
  Group}\label{sec:environment}
The effect of stripping that separates the luminous and non-luminous
satellites also extends to a marked difference between satellite and
field halos within the Local Group, and differences in the
distribution of luminous and dark halos beyond the virial radius of
its two main galaxies.

In Fig.~\ref{fig:matter-density} we show the $z=0$ distributions of
halos above $10^7\Ms$ within 2 Mpc to the centre of one of our Local
Group volumes, and within 400 kpc of one of the Milky-Way like
halos. It can be seen that low-mass luminous halos are predominantly
found close to the larger galaxies, where stripping has lead to a
significant reduction in mass.

While ram pressure stripping can quench star formation and introduce
environmental dependencies on the properties of low mass galaxies, gas
is not responsible for the environmental dependence of the luminous
fraction. It is primarily driven by the fact that, as a result of
tidal stripping, halos of a given mass or $\vm$ today that are closer
to the central galaxies typically had higher masses or $\vm$ in the
early universe, and therefore a higher probability for star formation.

In Fig.~\ref{fig:dark-fraction-distance} we show the fraction of halos
of a given mass or $\vm$ today that are populated by galaxies, as a
function of distance to the Milky-Way-like galaxy. It can be seen that
while all satellites with $M>10^9\Ms$ or $\vm > 20$~kms$^{-1}$, as
well as a large fraction of lower mass halos, are populated inside 200
kpc, the fraction decreases with increasing distance (an increase at
$\sim 800$kpc can be attributed to the satellites of the second
massive halo in the simulated Local Group, i.e. the halo of ``M31'').

This result implies that there are fewer isolated dwarf galaxies in
the Local Group than a simple extrapolation based on $\vm$ or mass
from the Milky Way and M31 satellites would suggest.

\section{Discussion and Summary}~\label{sec:summary} 

We have analysed a set of hydrodynamic and dark matter only (DMO)
simulations of the Local Group in order to study how galaxies populate
low mass halos in the presence of reionization. We have shown that
reionization greatly reduces the number of dwarf galaxies that form,
and that even at fixed mass or $\vm$ today, the subset of halos that
host the Local Group's dwarf galaxies differ significantly from the
total halo population.

The differences between the luminous and non-luminous halos result
from a combination of early-time and late-time effects. Halos of mass
below $\sim3\times10^9\Ms$ or $\vm$ below $\sim25$~kms$^{-1}$, that
host the majority of the Local Group dwarf galaxies, assembled their
mass significantly earlier than typical dark matter halos of the same
present-day mass or $\vm$. Since halos that form earlier are more
concentrated, and also because more concentrated halos are more
resistant to photo-evaporation, those halos that contain stars have
significantly higher $\vm$~--~mass ratios.

The late-time evolution of halos, and in particular tidal stripping,
introduces further biases between the luminous and non-luminous
halos. The halos of satellite galaxies with mass and $\vm$ values
similar to those of the observed dwarf spheroidals experienced much
stronger tidal stripping than comparable non-luminous satellites. The
Milky Way or M31 satellite galaxies typically formed in halos that
were more than twice as massive and had significantly higher $\vm$
values prior to infall. They also had earlier infall redshifts, higher
infall velocities, and followed more radial orbits than typical dark
matter halos.

Within the Local Group, isolated halos are much less likely to host
galaxies than satellites of comparable mass or $\vm$ today. In
particular, present-day field dwarf galaxies that share the
characteristics of dwarf spheroidals are more likely to be ``escaped''
satellites than was assumed from earlier DMO studies.

The biases between the luminous and non-luminous halos described here
have to be taken into account whenever the properties of halos or of
the underlying cosmology are to be measured, but only those halos
that contain galaxies can be observed.

It is worth noting that our simulations still have several
limitations. They assume that reionization is uniform and do not
account for local sources of ionization. Radiative transfer is not
included, cooling rates are computed in the optically thin limit, star
formation is modelled in a stochastic way, and we did not include star
formation in mini halos powered by H$_2$ cooling. While we have shown
that the fraction of halos populated by galaxies is numerically
converged, it may still depend on model parameters.

With these limitations in mind, we would like to emphasise the
importance of the connection between the observable dwarf galaxies and
the underlying population of dark matter halos. As noted in the
beginning, the Local Group dwarf galaxies provide the best window for
studying the nature of dark matter on small scales. However, our
results suggest that because reionization leaves most halos empty,
Local Group dwarf galaxies today {\it typically} live in halos that
are highly {\it atypical}.

\section*{Acknowledgements} \label{lastpage} We are indebted to
Dr. Lydia~Heck who ensures that the computers run smoothly at the
ICC. This work was supported by the Science and Technology Facilities
Council [grant number ST/F001166/1 and RF040218], the European
Research Council under the European Union's Seventh Framework
Programme (FP7/2007-2013) / ERC Grant agreement
278594-GasAroundGalaxies, the National Science Foundation under Grant
No. PHYS-1066293, the Interuniversity Attraction Poles Programme of
the Belgian Science Policy Office [AP P7/08 CHARM] and the hospitality
of the Aspen Center for Physics. T.~S. acknowledges the Marie-Curie
ITN CosmoComp. C.~S.~F. acknowledges an ERC Advanced Investigator
Grant COSMIWAY. This work used the DiRAC Data Centric system at Durham
University, operated by the Institute for Computational Cosmology on
behalf of the STFC DiRAC HPC Facility (www.dirac.ac.uk), and resources
provided by WestGrid (www.westgrid.ca) and Compute Canada / Calcul
Canada (www.computecanada.ca). The DiRAC system is funded by BIS
National E-infrastructure capital grant ST/K00042X/1, STFC capital
grant ST/H008519/1, STFC DiRAC Operations grant ST/K003267/1, and
Durham University. DiRAC is part of the National E-Infrastructure.

\bibliographystyle{mn2e}
\bibliography{dark_paper}

\end{document}